\documentclass[12pt,
aps,
floatfix,
 pra,reprint,
 superscriptaddress,
longbibliography
]{revtex4-1}
\usepackage{dcolumn}
\usepackage{bm}
\usepackage%
{graphicx}
\usepackage{xfrac}
\usepackage{extdash}
\usepackage{epsfig}
\usepackage{isomath}
\usepackage{textcomp}
\usepackage{bm}
\usepackage[english]{babel}
\usepackage{todonotes}
\usepackage{color,soul}
\usepackage{amsfonts}
\usepackage[columnwise]{lineno}

\usepackage{times}

\DeclareFontFamily{U}{euc}{}
\DeclareFontShape{U}{euc}{m}{n}{<-6>eurm5<6-8>eurm7<8->eurm10}{}%
\DeclareSymbolFont{AMSc}{U}{euc}{m}{n} 
\DeclareMathSymbol{\umu}{\mathord}{AMSc}{"16} 

\renewcommand{\vec}[1]{\boldsymbol{#1}}
\newcommand{\ensuretext}[1]{\ensuremath{\text{#1}}}

\newcommand{\unit}[1]{\ensuretext{\textrm{\,}}\ensuremath{\mathrm{#1}}}

\newcommand{\Mum}{\ensuremath{\umu}\ensuremath{\mathrm{m}}}
\newcommand{\mum}{\textrm{\,\ensuremath{\mathrm{\Mum}}}}
\newcommand{\eqref}[1]{(\ref{#1})}

\begin{document}

\title{Visualizing Plasmons and Ultrafast Kinetic Instabilities in Laser-Driven Solids using X-ray Scattering} 

\author{Paweł Ordyna}
\email{p.ordyna@hzdr.de}
\affiliation{Helmholtz-Zentrum Dresden-Rossendorf, Bautzner Landstra\ss e 400, 01328, Dresden, Germany}
\affiliation{Technical University Dresden, 01069 Dresden, Germany}
\author{Carsten B\"ahtz}
\affiliation{Helmholtz-Zentrum Dresden-Rossendorf, Bautzner Landstra\ss e 400, 01328, Dresden, Germany}
\author{Erik Brambrink}
\affiliation{European XFEL, Holzkoppel 4, 22869 Schenefeld, Germany}
\author{Michael Bussmann}
\affiliation{Helmholtz-Zentrum Dresden-Rossendorf, Bautzner Landstra\ss e 400, 01328, Dresden, Germany}
\author{Alejandro Laso Garcia}
\affiliation{Helmholtz-Zentrum Dresden-Rossendorf, Bautzner Landstra\ss e 400, 01328, Dresden, Germany}
\author{Marco Garten}
\altaffiliation[Present address: ]{Lawrence Berkeley National Laboratory,  1 Cyclotron Rd, Berkeley, CA 94720, USA}
\affiliation{Helmholtz-Zentrum Dresden-Rossendorf, Bautzner Landstra\ss e 400, 01328, Dresden, Germany}
\author{Lennart Gaus}
\affiliation{Helmholtz-Zentrum Dresden-Rossendorf, Bautzner Landstra\ss e 400, 01328, Dresden, Germany}
\affiliation{Technical University Dresden, 01069 Dresden, Germany}
\author{Sebastian G\"ode}
\affiliation{European XFEL, Holzkoppel 4, 22869 Schenefeld, Germany}
\author{J\"org Grenzer}
\affiliation{Helmholtz-Zentrum Dresden-Rossendorf, Bautzner Landstra\ss e 400, 01328, Dresden, Germany}
\author{Christian Gutt}
\affiliation{Universit\"at Siegen, Walter-Flex Stra\ss e 3, 57072 Siegen, Germany}
\author{Hauke H\"oppner}
\affiliation{Helmholtz-Zentrum Dresden-Rossendorf, Bautzner Landstra\ss e 400, 01328, Dresden, Germany}
\author{Lingen Huang}
\affiliation{Helmholtz-Zentrum Dresden-Rossendorf, Bautzner Landstra\ss e 400, 01328, Dresden, Germany}
\author{Oliver Humphries}
\affiliation{European XFEL, Holzkoppel 4, 22869 Schenefeld, Germany}
\author{Brian Edward Marré}
\affiliation{Helmholtz-Zentrum Dresden-Rossendorf, Bautzner Landstra\ss e 400, 01328, Dresden, Germany}
\affiliation{Technical University Dresden, 01069 Dresden, Germany}
\author{Josefine Metzkes-Ng}
\affiliation{Helmholtz-Zentrum Dresden-Rossendorf, Bautzner Landstra\ss e 400, 01328, Dresden, Germany}
\author{Motoaki Nakatsutsumi}
\affiliation{European XFEL, Holzkoppel 4, 22869 Schenefeld, Germany}
\author{\"Ozg\"ul \"Ozt\"urk}
\affiliation{Universit\"at Siegen, Walter-Flex Stra\ss e 3, 57072 Siegen, Germany}
\author{Xiayun Pan}
\affiliation{Helmholtz-Zentrum Dresden-Rossendorf, Bautzner Landstra\ss e 400, 01328, Dresden, Germany}
\affiliation{Technical University Dresden, 01069 Dresden, Germany}
\author{Franziska Paschke-Br\"uhl}
\affiliation{Helmholtz-Zentrum Dresden-Rossendorf, Bautzner Landstra\ss e 400, 01328, Dresden, Germany}
\author{Alexander Pelka}
\affiliation{Helmholtz-Zentrum Dresden-Rossendorf, Bautzner Landstra\ss e 400, 01328, Dresden, Germany}
\author{Irene Prencipe}
\affiliation{Helmholtz-Zentrum Dresden-Rossendorf, Bautzner Landstra\ss e 400, 01328, Dresden, Germany}
\author{Thomas R Preston}
\affiliation{European XFEL, Holzkoppel 4, 22869 Schenefeld, Germany}
\author{Lisa Randolph}
\thanks{European XFEL is the present adress}
\affiliation{Universit\"at Siegen, Walter-Flex Stra\ss e 3, 57072 Siegen, Germany}
\affiliation{European XFEL, Holzkoppel 4, 22869 Schenefeld, Germany}
\author{Hans-Peter Schlenvoigt}
\affiliation{Helmholtz-Zentrum Dresden-Rossendorf, Bautzner Landstra\ss e 400, 01328, Dresden, Germany}
\author{Jan-Patrick Schwinkendorf}
\affiliation{Helmholtz-Zentrum Dresden-Rossendorf, Bautzner Landstra\ss e 400, 01328, Dresden, Germany}
\affiliation{European XFEL, Holzkoppel 4, 22869 Schenefeld, Germany}
\author{Michal \v{S}m\'{i}d}
\affiliation{Helmholtz-Zentrum Dresden-Rossendorf, Bautzner Landstra\ss e 400, 01328, Dresden, Germany}
\author{Sebastian Starke}
\affiliation{Helmholtz-Zentrum Dresden-Rossendorf, Bautzner Landstra\ss e 400, 01328, Dresden, Germany}
\author{Radka Stefanikova}
\affiliation{Helmholtz-Zentrum Dresden-Rossendorf, Bautzner Landstra\ss e 400, 01328, Dresden, Germany}
\affiliation{Technical University Dresden, 01069 Dresden, Germany}
\author{Erik Thiessenhusen}
\affiliation{Helmholtz-Zentrum Dresden-Rossendorf, Bautzner Landstra\ss e 400, 01328, Dresden, Germany}
\affiliation{Technical University Dresden, 01069 Dresden, Germany}
\author{Toma Toncian}
\affiliation{Helmholtz-Zentrum Dresden-Rossendorf, Bautzner Landstra\ss e 400, 01328, Dresden, Germany}
\author{Karl Zeil}
\affiliation{Helmholtz-Zentrum Dresden-Rossendorf, Bautzner Landstra\ss e 400, 01328, Dresden, Germany}
\author{Ulrich Schramm}
\author{Thomas E. Cowan}
\affiliation{Helmholtz-Zentrum Dresden-Rossendorf, Bautzner Landstra\ss e 400, 01328, Dresden, Germany}
\affiliation{Technical University Dresden, 01069 Dresden, Germany}
\author{Thomas Kluge$^{1,}$}
\email{t.kluge@hzdr.de}

\date{\today}

\begin{abstract}
Ultra-intense lasers that ionize and accelerate electrons in solids to near the speed of light can lead to kinetic instabilities that alter the laser absorption and subsequent electron transport, isochoric heating, and ion acceleration. 
These instabilities can be difficult to characterize, but a novel approach using X-ray scattering at keV photon energies allows for their visualization with femtosecond temporal resolution on the few nanometer mesoscale. 
Our experiments on laser-driven flat silicon membranes show the development of structure with a dominant scale of $~60\unit{nm}$ in the plane of the laser axis and laser polarization, and $~95\unit{nm}$ in the vertical direction with a growth rate faster than $0.1/\mathrm{fs}$. 
Combining the XFEL experiments with simulations provides a complete picture of the structural evolution of ultra-fast laser-induced plasma density development, indicating the excitation of plasmons and a filamentation instability. 
Particle-in-cell simulations confirm that these signals are due to an oblique two-stream filamentation instability. 
These findings provide new insight into ultra-fast instability and heating processes in solids under extreme conditions at the nanometer level with possible implications for laser particle acceleration, inertial confinement fusion, and laboratory astrophysics. 
\end{abstract}
\maketitle 

Visualizing, understanding and controlling laser absorption, isochoric heating, particle acceleration, and other relativistic non-linear physics that occur at the interaction of powerful lasers with solids is important for applications ranging from next-generation laser ion accelerators (LIA) for medical use\cite{Kroll2022} to high-energy density physics including laboratory astrophysics\cite{Albert2021} and inertial confinement fusion\cite{Craxton2015,Edwards2015}. 
Only recently, (proton) fast ignition for inertial confinement fusion has gained renewed interest as a viable path towards commercialization of Inertial Fusion Energy\cite{Zylstra2022,*Wilks2022,*Obst-Huebl2022} after the breakthrough fusion ignition achievements at the National Ignition Facility (NIF)\cite{Abu-Shawareb2022,Bishop2022}. \\
Of special relevance is the understanding and control of plasma instabilities. 
For example, compression and ignition of fusion targets in indirect-drive experiments carried out e.g. at the NIF nanosecond laser rely on the conversion of the laser energy into a homogeneous radiation field by laser-self-generated grating structures at the hohlraum inner walls\cite{Glenzer2010}. 
Here we focus on ultrafast few femtosecond relativistic instabilities that are important e.g. in fusion fast ignition scenarios (FIS) \cite{Tabak1994,Roth2001,Campbell2021} that could potentially allow for a much better efficiency. 
Small fluctuations in the radiation pressure on the pellet surface or in the particle heater pulse would otherwise drive instabilities there, inhibiting maximum compression or heating \cite{Mahdavi2021,Badziak2023}. 

Theories for instabilities in relativistic high-intensity laser interaction with solids fall into two categories: (i) hydrodynamic instabilities growing at interfaces between two fluid-like plasma or photon ensembles, or (ii) kinetic instabilities that occur e.g. when one plasma streams through the other. 
Whether the one or the other dominate depends on the detailed laser and solid properties. 
For example, in solids with a structured surface, or driven by lasers with a shallow rising edge, laser absorption to relativistic electron currents reaches up to 100\%, emphasizing the kinetic streaming instabilities at the front surface \cite{Bret2004,Metzkes2014a} or at the rear of the target\cite{Gode2017,Scott2017,Huang2017}, e.g. two-stream instability (TSI), Weibel instability (WI), or filamentation instability (FI). 
On the other hand, strong hydrodynamic Rayleigh-Taylor-like instabilities (RTI) following two-plasmon decay or parametric instabilities at the front of plasmas can be dominant for materials consisting of light ions, or driven by ultra-short high-contrast laser pulses, and can break up the laser to electron coupling and inhibit streaming instabilities\cite{Macchi2002,Palmer2012,Kluge2015,Sgattoni2015}. 

The physics of these fast few femtosecond, few nanometer plasma instability dynamics and merging to the micron-scale after a few picoseconds in high-intensity laser driven solids is one of the large unsolved issues in high-intensity laser plasma science, but its direct observation has previously not been possible because of the small time and few nanometer length scales involved. 
Microscopic interpretations were therefore primarily based on simulations and indirect measurements, e.g. via optical microscopy~\cite{Sokolowski-Tinten1998}, interferometry\cite{Geindre1994, *Bocoum2015}, spectroscopy\cite{Malvache2013}, or radiography\cite{Quinn2012,Gode2017,Scott2017}. 
Here we demonstrate experimentally that such instabilities indeed exist in the hot solid density plasmas, quantify the strength, and give limits to the growth rate. 

Recent advances in the time-resolved diffraction, based on ultra-fast X-ray pulses from XFELs now enable us to investigate laser produced plasmas with nanometer spatial and femtosecond temporal resolution~\cite{Fletcher2015,Gorkhover2016,Kluge2017,Kluge2018,Mo2019,Gaus2021}. 
The novelty of the present work is not the 
technique (which is indeed very similar to the setup used in \cite{Kluge2017,Gaus2021}) nor the specific 
physical scenario probed with the technique. 
In fact, since the scenario is of high relevance for FIS or LIA, there exists an exhaustive theoretical work (e.g. \cite{Davidson1972,Tatarakis2003,Bret2004,Bret2005,Bret2005a,Cottrill2008,Bret2010,bret2014}). 
The novelty of the work is rather that it is the first direct observation of laser-induced instabilities in ultra-intense laser-solid interaction with few tens of nanometer and femtosecond spatial and temporal resolution. 
Most of the previous studies focused on the irradiation and ablation of gratings or other pre-structured targets. 
Here we open a new window for observations of structural dynamics induced by the laser interaction with solids, not relying on prefabricated structures anymore, which allows for benchmarking of the laser absorption, laser generated electron currents, as well as thermalization and diffusion processes. 
Specifically, by variation of the laser and target parameters one can map out the instability spectrum of the fastest growing mode, i.e. the growth rate as a function of the instability wave vector, and get access to the dielectric tensor. 
This in turn is determined by the particle momentum distribution function, which together with its temporal evolution could thus be measured in pump probe SAXS experiments.

Here, we demonstrate the feasibility of this approach by showing the experimental realization of probing the predicted nanoscopic instability growth over few femtoseconds. 
This proved difficult in the past, as the world's most powerful optical drive lasers need to be combined with the most advanced X-ray sources, and experiments faced fundamental challenges such as the parasitic bremsstrahlung generation and self emission of the warm or hot dense plasma that can outshine the signal. 

A tightly connected phenomenon that occurs when laser accelerated electron beams propagate through dense plasmas is the generation of bulk plasmons (BP). 
For example, BPs in solids can anomalously heat the bulk\cite{Sherlock2014} or decay into surface plasmons \cite{Kluge2015} that due to their unique properties are a building block with several applications\cite{Macchi2018}, e.g. accelerating electrons and ions\cite{Fedeli2016}, generating XUV radiation, and isocorically heating the surface. \\

\begin{figure}
\centering
  \includegraphics[width=\linewidth]{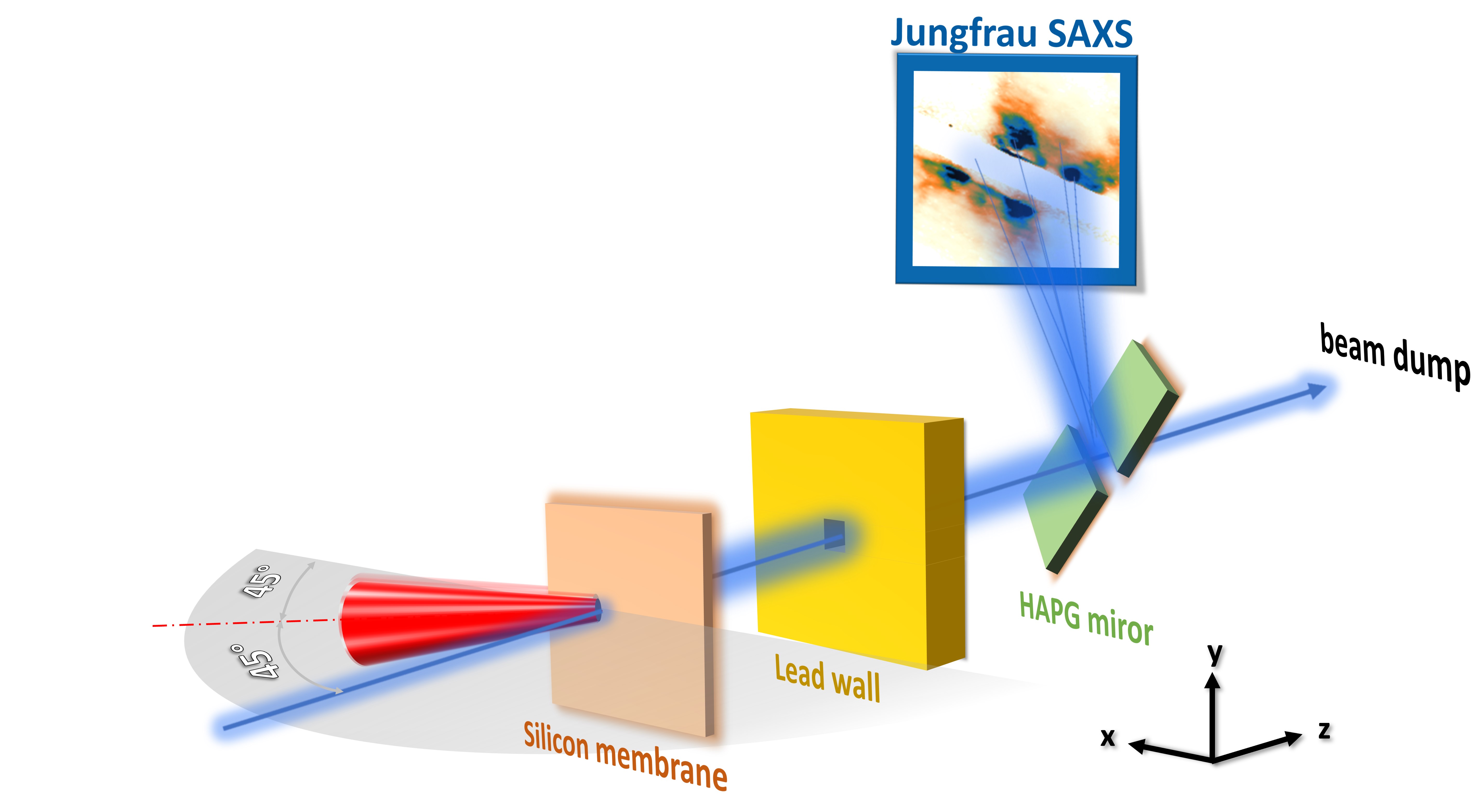}
\caption{Experimental setup, not to scale. The ReLaX UHI laser (red) is focused onto the Silicon membrane target under $45^\circ$ in p-polarization, the XFEL (blue) is probing the plasma density under target normal direction. The Jungfrau CCD detector records the SAXS image reflected from the HAPG chromatic mirror.}
  \label{fig:setup}
\end{figure}

\begin{figure*}
\centering
  \includegraphics[width=0.75\linewidth]{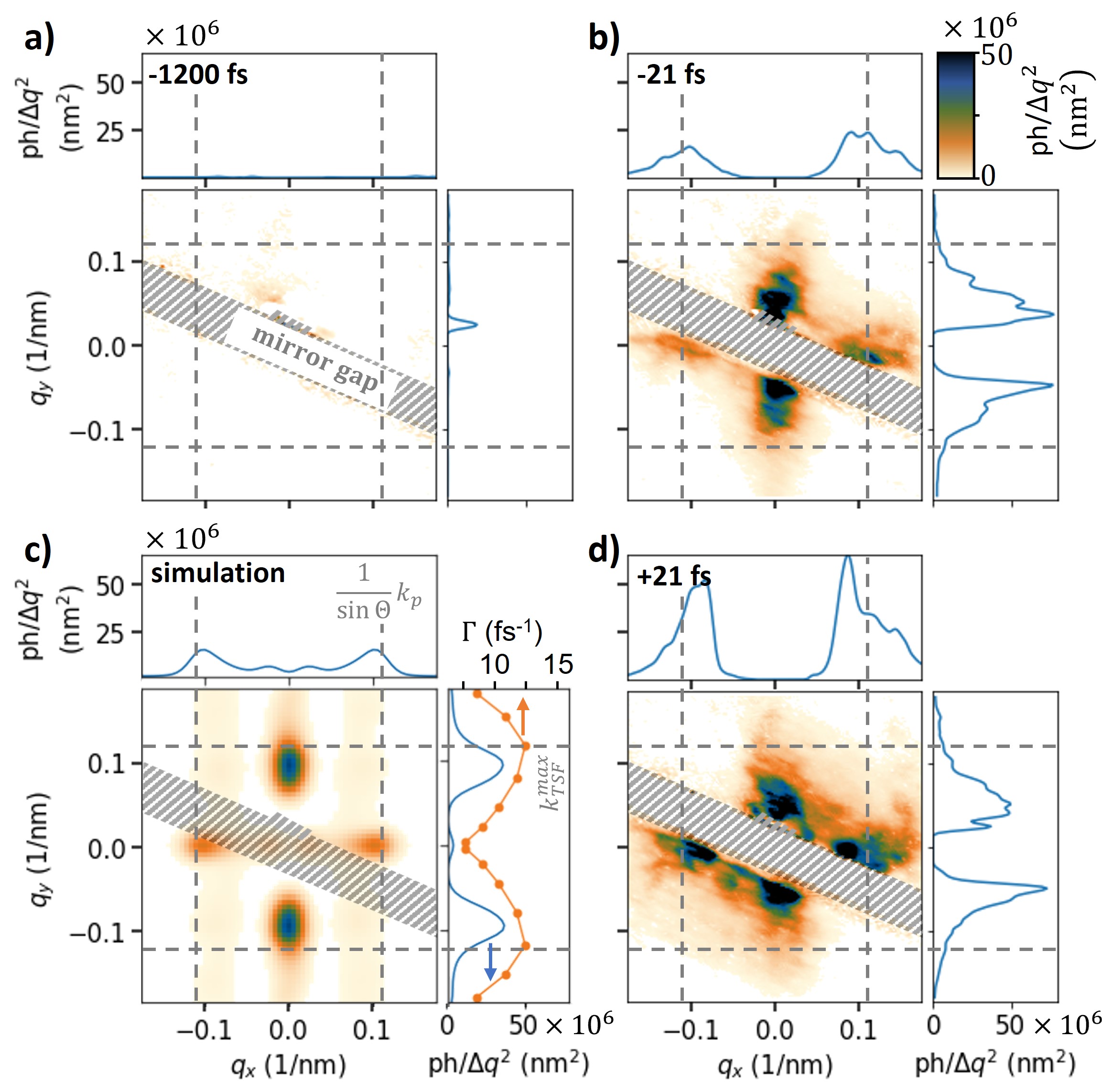}
\caption{The background subtracted (as determined by an XFEL-only pre-shot train) and de-noised (using a neural-network (NN) based algorithm) scattering patterns recorded by the SAXS detector for three exemplary runs (corrected for HAPG mirror reflectivity and geometry, see methods), the respective probe time delay is indicated in the top panels. The absolute zero delay was defined as the mean value between the two highest signal shots, but was not measured, the relative timing error is below $15\unit{fs}$ (see methods). The lineouts were taken through $\vec{q}=0$, averaged over $\pm 0.01/\mathrm{nm}$ around $q=0$. A simulated scattering pattern is shown for $t=0$. The dashed lines show the analytic theoretical expectation from numerically solving the dispersion relation (y-direction, and bulk plasmons (x-direction, Eqn.~\eqref{eqn:plasmons}).  }
  \label{fig:experiment1}
\end{figure*}

With the recent completion\cite{lasogarcia2021} and commissioning of the Helmholtz International Beamline for Extreme Fields (HIBEF\cite{HIBEFConsortium}) at the European XFEL\cite{Tschentscher2013,Zastrau2021}, the quest for visualising few-femtosecond, few-nanometer scale non-linear plasma dynamics in ultra-short pulse ultra-high intensity (UHI) laser interaction with solids has begun. \\
One primary goal of HIBEF as well as this work is to provide novel experimental benchmarks for high energy density science, including the measurement and characterization of kinetic instabilities in order to validate codes and theoretical models, and to optimize laser absorption and successive processes for above-mentioned applications. 
HIBEF combines a short-pulse Titanium:Sapphire UHI laser (ReLaX) with the European XFEL beam. 
The ReLaX laser reaches its highest intensity of $5\cdot 10^{20}\unit{W/cm}^2$ when it is focused down to a $4.7\unit{\mum}$ (FWHM) focal spot and compressed to $\tau_L = 30\unit{fs}$ pulse duration, exceeding the laser intensity at other XFELs by approx. an order of magnitude\cite{lasogarcia2021}. 
One enabling technology that was developed by our group for HIBEF is a SAXS detection system designed primarily to suppress radiation background. 
Parasitic radiation is not only suppressed by carefully designed lead shielding inside the chamber, at the chamber exit window, and around the X-ray camera. 
Crucially, we developed a passive radiation background suppression system, i.e. a set of HAPG X-ray crystals\cite{Smid2020} between the solid target and the SAXS detector. 
They act as a chromatic X-ray mirror that separates the signal $8\unit{keV}$ X-ray energy from the background. 
We thereby effectively suppressed almost any X-ray background and plasma self-emission which at ultra-relativistic laser intensities above $10^{20}\unit{W/cm}^2$ would otherwise outshine the signal. 
Additionally, it is quite likely that in earlier studies employing sub-relativistic laser intensity the structure development was not sufficiently strong to be detected. 
To our knowledge, this is the first study combining an UHI laser, flat solid targets and an XFEL probe, enabling us for the first time to measure the signal generated by relativistic instabilities driven by a UHI short-pulse laser. 

We measured time-resolved SAXS patterns with sufficient momentum transfer range that provides, with the current setup (Fig.~\ref{fig:setup}), sensitivity for correlation structures up to $\cong 100\unit{nm}$. 
Larger correlations are not accessible due to a gap in the HAPG mirror that lets pass the unscattered XFEL beam towards a beam dump. 
Since the detector covers only the small angle $q$-range, the signal is essentially given by the time integration of the Fourier transform absolute square of the \emph{time retarded} electron density integrated along the XFEL beam direction
\begin{equation}
    I(\vec{q})\propto \int_t \left|\mathcal{F}_{\vec{r}} (E_X(\vec{r},t)\tilde n_e(\vec{r},t))\right|^2 dt \ ,
    \label{eqn:FT}
\end{equation}
where $\tilde n_e(\vec{r},t) = \int_z n_e(x,y,z,t'=t+z/c) dz$ is the time retarded electron density projection and $E_X$ is the XFEL electric field amplitude. 

Fig.~\ref{fig:experiment1} shows the measured scattering patterns as a function of probe delay from probing flat $2\mum$ thin silicon (Si) membranes irradiated by the ReLaX UHI laser at maximum intensity under $45^\circ$ angle of incidence and p-polarization. 
Each main shot was accompanied by a pre-shot and a post-shot XFEL-only pulse train on the same spot (for the pre-shots the XFEL transmission was reduced by a factor of $6\cdot 10^{-4}$, in order to protect the target from X-ray damage). 
This enabled us to verify the cold membrane quality, and to subtract parasitic signal in the background.
The scattering patterns show dominant signal around the ReLaX laser peak arrival time along the vertical direction (momentum transfer along the laser magnetic field direction) and horizontal direction ( momentum transfer in the plane of the laser axis, the laser electric field vector, and the target normal). 

In Fig.~\ref{fig:experiment2} we show the integrated number of photons recorded along the horizontal and vertical direction for all the 6 data shots that we took for this study. 
Note that up to $-1\unit{ps}$ no significant scattering was measurable, as expected from a flat membrane. 
At $t=-(31\pm 13)\unit{fs}$ the signals are still consistent with zero signal within a $2\sigma$ confidence interval. 
A large scattering signal then sharply occurs at $-(21\pm 13)\unit{fs}$, remaining high for approximately the laser pulse duration. 
This is an ultra-fast temporal growth of more than two orders of magnitude within $10-30\unit{fs}$, which shows that the measurements have indeed happened during or shortly after the UHI laser irradiation. 

These scattering signals are indicative for a growth of correlated plasma electron density modulations on the timescale of only a few femtoseconds. 
Simulations described in the next section (Fig.~\ref{fig:simulation}) confirm these findings. 
To identify the dominant instability mode, we solve the dispersion relation numerically. 
For our case we find the fastest growing mode is the oblique/two-stream filamentation instability (TSFI) in the front of the foil.
The filamentation part is responsible for the signal in vertical direction, while the horizontal signal is originating from scattering on plasmons excited by the laser accelerated electrons traversing the bulk and being subjected to the two-stream instability. 
A synthetic, forward-calculated scattering image showing the two scattering features from our simulations is shown in Fig.~\ref{fig:experiment1}c. 

\section*{Discussion of the results}
A cross-like pattern in Fourier space corresponds to a mesh-like pattern in real space. 
This means that the experimental data directly confirms without any further assumptions that the flat membrane must have developed a corresponding mesh-like electron density pattern in the electron density projected along the XFEL direction according to Eqn.~\eqref{eqn:FT} as has long been expected from simulations\cite{Bret2010,Kluge2014}. 
This is the first \emph{direct} measurement of plasma instabilities and plasmons in high-intensity laser-driven solids during or shortly after the irradiation on few nanometer spatial scale. 

To answer the question of the origin of the scattering pattern, we turn to possible instabilities that are known to generate a grating-like pattern. 
These include, for example, RTI, WI, or FI, but we could also consider a combination of different waves and instabilities for the different orientations. 
The situation is further complicated by the high sensitivity of growth rates e.g. to collisions\cite{Cottrill2008} and the momentum distribution of the beam and bulk electrons\cite{Bret2010}. 
In fact, since there exists no previous direct measurement of the solid density plasma break up during the ultra-short laser pulse irradiation in literature, we have to resort to simulations of the UHI laser interaction with the silicon membrane in order to identify and describe quantitatively the plasma dynamics at play. 

\begin{figure}
\centering
  \includegraphics[width=\linewidth]{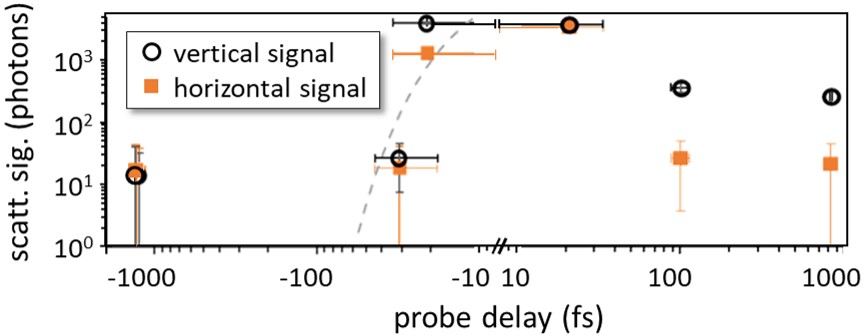}
\caption{Integrated background subtracted scattering signal strength as a function of probe delay. The projected signal around the peaks was integrated over the scattering signal above the noise level in the respective direction. The probe delay was extracted from the pulse arrival monitor, the relative timing uncertainty is below $15\unit{fs}$ for all shots (absolute timing as in Fig.~\ref{fig:experiment1}). }
  \label{fig:experiment2}
\end{figure}
\begin{figure*}
\centering
  \includegraphics[width=\linewidth]{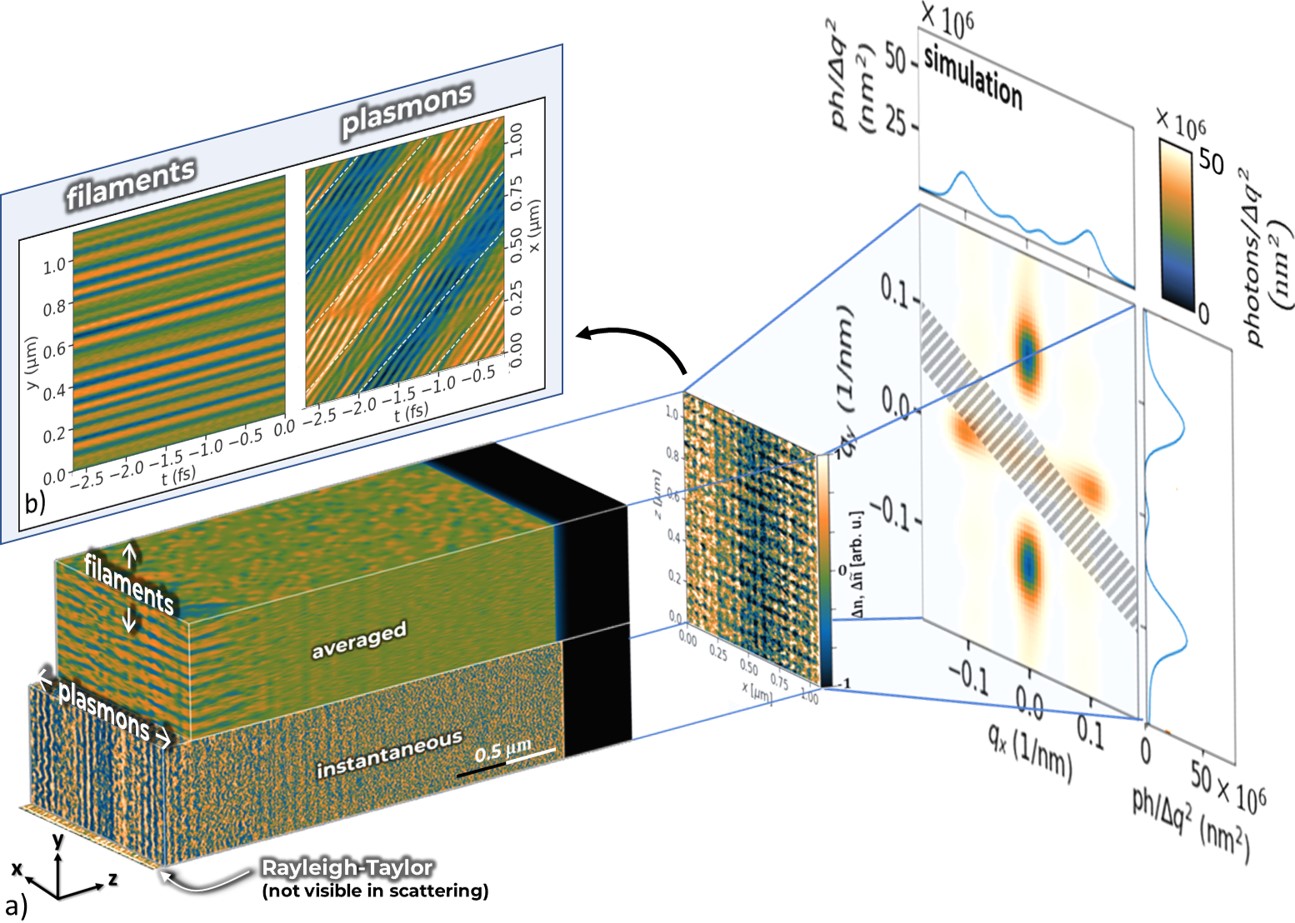}
\caption{Simulation: \textbf{a)} \textbf{Left:} Instantaneous and averaged (over a plasma wavelength of $40\unit{nm}$) electron density deviation from the average target density close to $t=0$. The superposition of both creates the mesh-like pattern seen in $\tilde n_e$ \textbf{(center)}. The mesh generates a cross-like pattern in the synthetic scattering signal $\propto \mathfrak{F}(\tilde n_e)$ \textbf{(right)}. \textbf{b)} Time evolution of $\tilde n_e$ averaged along $x$ (left) and $y$ (right) show that the filamentation pattern is static while the plasmon features are moving with close to $\sqrt{2}c$. 
}
  \label{fig:simulation}
\end{figure*}

In Fig.~\ref{fig:simulation} we summarize the results of our simulation, which suggests that the scattering signal that corresponds to a mesh-like projected density profile is in fact generated by two rather independent structures that each generate a comb-like density, rotated $90^{\circ}$ w.r.t each other. \\
First, we find a rapidly growing instability mode in the \emph{vertical direction} (i.e. a density comb in the y-direction with horizontal lines along the x-direction). The density modulations extend inside the target from the front surface with a filament distance of $\sim 63\unit{nm}$ (i.e. $k_y^{PIC}\sim 0.1/nm$) and are witnessed by corresponding magnetic field filaments. 
The structure grows within several femtoseconds during the laser irradiation (growth rate $\Gamma_{y}^{sim}\cong 0.1\unit{fs}^{-1}$) and is spatially static, see Fig.~\ref{fig:simulation}b. 
The lower bound for the growth rate extracted from the experiment (dashed line Fig.~\ref{fig:experiment2}, $1\sigma$ confidence interval) is with $0.09/\mathrm{fs}$ in good agreement with the simulation, while the spatial scale corresponding to the experimentally observed scattering wave vector is somewhat larger than in the simulation. \\
The simulation suggests that the vertical instability is due to the inhomogeneous current distribution, since it is spatially aligned along the laser generated electron current. 
To identify the exact instability mode one has to compute the dielectric tensor and find the solution to the dispersion relation that maximizes the imaginary part of the wave frequency. 
Because analytic theories rely on specific idealized electron distribution functions, we cannot directly use them here. 
For example, it is not obvious how to define the fast forward and the fast and bulk return currents, as they continuously merge into another\cite{Sherlock2016}. 
Rather, we extract the full electron distribution function from the simulation and solve both the integrals in the dielectric tensor and the dispersion relation numerically (i.e. Eqn.~6, and Eqns.~16 and 17 in \cite{Bret2004} for the Weibel and two-stream filamentation (TSF) branch of the dispersion relation, respectively), assuming the ions are at rest and neglecting magnetic fields. 
As one might expect from the analytic treatment in the idealized two-stream case, the pure filamentation and Weibel-like modes on the TSF-branch are lost due to the effects of transverse beam and plasma temperature (or more generally: broad energy distribution), respectively. 
Instable areas are recovered both on the Weibel branch as well as oblique modes on the TSF branch. 
Yet, in our case the growth rates observed in the latter largely dominate over the former, so that the system is dominated by the oblique modes, also called electromagnetic beam-plasma instability\cite{Califano1998} or two-stream filamentation instability\cite{Bret2004}. 
While for the cold case the maximum growth rate of the TSFI is found at $k^{max}=\infty$, temperature effects stabilize small wavelengths by preventing pinching to small radii\cite{Silva2002}. 
Numerically solving the dispersion relation we find the growth rate is maximized at $k_y^{max}=0.12/nm$ with $\Gamma^{max}=12/fs$ (orange line in Fig.~\ref{fig:experiment2}). 
While this value of $k_y^{max}$ is close to $k_y^{PIC}$ in the simulation, the growth rate is much larger than $\Gamma_{y}^{sim}$. 
This would mean that the instability is in continuous saturation and in the simulation we rather measure the growth of the saturation limit. \\
In the \emph{horizontal direction} the integrated density from the simulation shows a dynamic electron plasmon wave structure moving with the phase velocity $v_{p,x} = c/\sin{\Theta}$ transverse to the XFEL, where $\Theta=\pi/4$ is the laser incidence angle w.r.t. the target normal. 
Along the laser direction electrons are accelerated which excite bulk plasmons as they propagate through the plasma, which then occur as a travelling periodic wave-like density comb with vertical lines in the density projection along the XFEL direction with wave vector 
\begin{equation}
    k_{p,x} = \frac{\omega_{pe}}{v_{p,x}} \approx 0.11 \frac{1}{\mathrm{nm}}
    \label{eqn:plasmons}
\end{equation}
corresponding to $\lambda_{p,x} = 56\unit{nm}$ ($\omega_{pe}\cong 47\unit{fs}^{-1}$ is the plasma frequency), see right panel in Fig.~\ref{fig:simulation}b. \\ 
It is important to point out that this dynamic feature can only be measured by means of scattering, since in shadowgraphic probing the projection of the \emph{time-integrated} density is measured, which would almost completely smear out for the travelling plasmons and probes longer than the plasmon period. 

A synthetic scattering pattern computed from the simulation over $2$ laser periods around the peak intensity arrival on target is shown in Fig.~\ref{fig:experiment1} and allows a direct comparison with the experiment. 
The general structures are in good agreement with the experiment, both quantitatively and qualitatively. 
In the horizontal direction the scattering peaks are nearly at the same position in simulation and experiment, while there is a slight mismatch in the vertical direction. 
Indirect measurements always had the problem that such small deviations could in principle be attributed to the complex processes involved in the measurement, in contrast here we have a clear indication for a different TSFI wavelength measured than observed in the simulation which is indicative for differences in the electron momentum distribution, i.e. the laser generated electron current and return current (e.g. beam energy, temperature, density). \\ 
Another important aspect is the intensity ratio between the vertical and horizontal signal. 
While the filaments exist over the full laser pulse duration and beyond, due to the comparatively slow magnetic diffusion time scales, the plasmonic signal essentially stops when the laser pulse is over. 
Hence, at later times we expect the vertical signal to be much larger than the horizontal one, by up to two orders of magnitude at $30\unit{fs}$ after the laser peak based on our simulation. 
The fact that we experimentally observe this strong difference between the vertical and horizontal signal strengths in deed only at the larger delays $t \geq 100\unit{fs}$ is therefore an additional indication for having probed the plasma around the laser peak. \\

In conclusion, combining the experiments, simulations and analytic estimates, we can draw a complete picture of the dominant plasma dynamics in the current experiment: As the relativistic laser accelerated forward electron current streams through the bulk return current transverse filaments are growing rapidly during the laser irradiation, and at the same time longitudinal plasmons are driven. 
Both, filaments and plasmons, add up to generate a mesh-like electron density pattern that is responsible for the measured cross-like scattering pattern. 
This is the first ultra-fast dynamic signal visualized from UHI laser-driven solids on the few nanometer scale, highlighting the great potential of SAXS for studying the early time of UHI-solid interaction dynamics.
We measured the spatial electron density correlations to few-nanometer, few-femtosecond precision, which is in reasonable qualitative agreement with our simulations that favour the TSFI. 
More comprehensive measurements of the growth rates and spatial scales will allow to refine and benchmark our simulations and overall knowledge of important key topics in relativistic plasma physics, including laser absorption, return current generation, instability growth and thermal stabilization, via the dispersion relation. 
For example, the growth rate dependency on the electron beam Lorentz factor $\gamma$ is distinctly different between WI ($\propto \gamma^{-1}$), FI ($\propto \gamma^{-1/2}$), and TSFI ($\propto \gamma^{-1/3}$), and the instability spectrum in $k$-space depends on the electron momentum distribution via the dielectric tensor. 

\section*{Data availability}
Data recorded for the experiment at the European XFEL are available at doi:10.22003/XFEL.EU-DATA-002854-00. The processed data and simulation data, as well as the scripts used to generate Figs.~\ref{fig:experiment1}-\ref{fig:simulation} are available at doi:1014278/rodare/2183.

\begin{acknowledgments}
We acknowledge European XFEL in Schenefeld, Germany, for provision of X-ray free-electron laser beamtime at HED (High Energy Density Science) HIBEF (Helmholtz International Beamline for Extreme Fields) SASE2 and would like to thank the staff for their assistance.
The authors are indebted to the HIBEF user consortium for the provision of instrumentation and staff that enabled this experiment. 
Christian Gutt acknowledges funding by DFG (GU 535/6-1). 
This work has also been supported by HIBEF (www.hibef.eu) and partially by the European Commission via H2020 Laserlab Europe V (PRISES) contract no. 871124, and by the German Federal Ministry of Education and Research (BMBF) under contract number 03Z1O511. 
This research used resources of the National Energy Research Scientific Computing Center (NERSC), a U.S. Department of Energy Office of Science User Facility located at Lawrence Berkeley National Laboratory, operated under Contract No. DE-AC02-05CH11231 using NERSC award FES-ERCAP0023216.
We would like to express our sincere gratitude to S.V.Rahul from European XFEL for his invaluable insights and fruitful discussions related to this publication. 
\end{acknowledgments}

\appendix
\section{Methods}
\subsection{Experimental procedure}
\subsubsection{ReLaX and XFEL properties} The Titanium:Sapphire high intensity short pulse laser ReLaX has a wavelength of $\lambda_L=0.8\unit{\mum}$. The measured focus size of the ReLaX laser was approx. $w_L=4.7\unit{\mum}$ FWHM. The laser pulse energy, duration and calculated intensity were $W_L\approx 2.8\unit{J}$, d$\tau_L=30\unit{fs}$, and $5\cdot 10^{20}\unit{W/cm}^2$ ($a_0=15$), respectively. \\ 
The XFEL beam was used in SASE configuration with a wavelength of $\lambda_X=0.15\unit{nm}$ ($=8\unit{keV}$). The focal spot was $w_X\approx 20\unit{\mum}$ FWHM, the XFEL energy, duration and calculated number of photons where $W_X \approx 1.5\unit{mJ}$, $\tau_X\approx 30\unit{fs}$, and $N_X\approx 1.2\cdot 10^{12}$ photons per bunch. The XFEL fundamental was dumped on an X-ray detector $~4\unit{m}$ downstream of the target.

\subsubsection{Synchronization} The XFEL probe time delay given in the figures is the relative timing measured with the HED optical encoding pulse arrival monitor 
 (PAM) to a precision of $12.9\unit{fs}$ w.r.t. the nominal zero delay set for all runs\cite{lasogarcia2021}. The nominal zero delay was not calibrated for the shots in this work, so that this is largely unknown due to drift and the jitter. We therefor give all times relative to the central time between the two highest yield shots. 
 
\subsubsection{SAXS signal processing} 
\begin{figure}[!h]
    \centering
    \includegraphics[width=\linewidth]{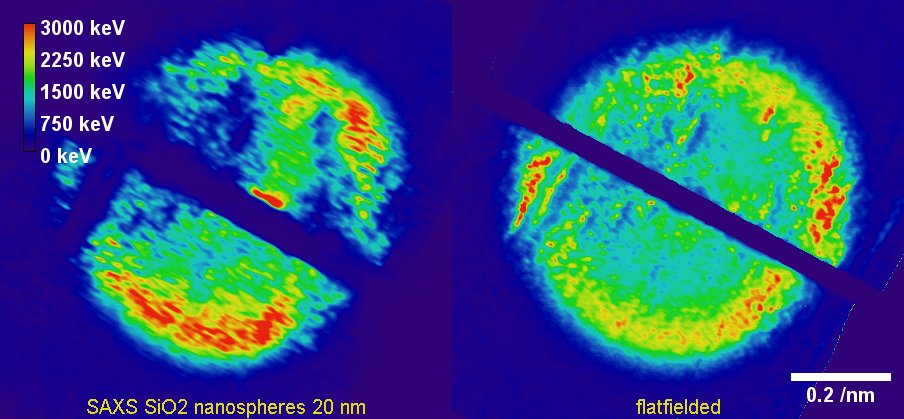}
    \caption{Scattering signal from SiO2 nanospheres ($20\unit{nm}$ diameter before (left) and after correction (right) as described in \cite{Smid2020}.}
    \label{fig:flatfield}
\end{figure}

THE SAXS signal is reflected by the HAPG mirror to the Jungfrau detector. The reflectivity of the mirror is approx. 0.2. As the reflectivity of the mirror is varying over its surface, it has to be corrected by a flat-field inferred from the scattering on a known substance, we employed SiO2 nanospheres target with particle diameter 20nm, as described in\cite{Smid2020}. The geometrical distortion of the signal was also corrected by a scheme described there, see Fig.~\ref{fig:flatfield}.
\\
\stepcounter{figure}
Then, the data was de-noised (cp. Fig.~\arabic{figure}) using a custom-made neural network trained on the experimental background added to synthetic data. The details of the algorithm available at [doi:1014278/rodare/xxxx] will be published in a separate paper. 
\begin{widetext}
    \centering
    \begin{minipage}{\linewidth}
            \includegraphics[width=\linewidth]{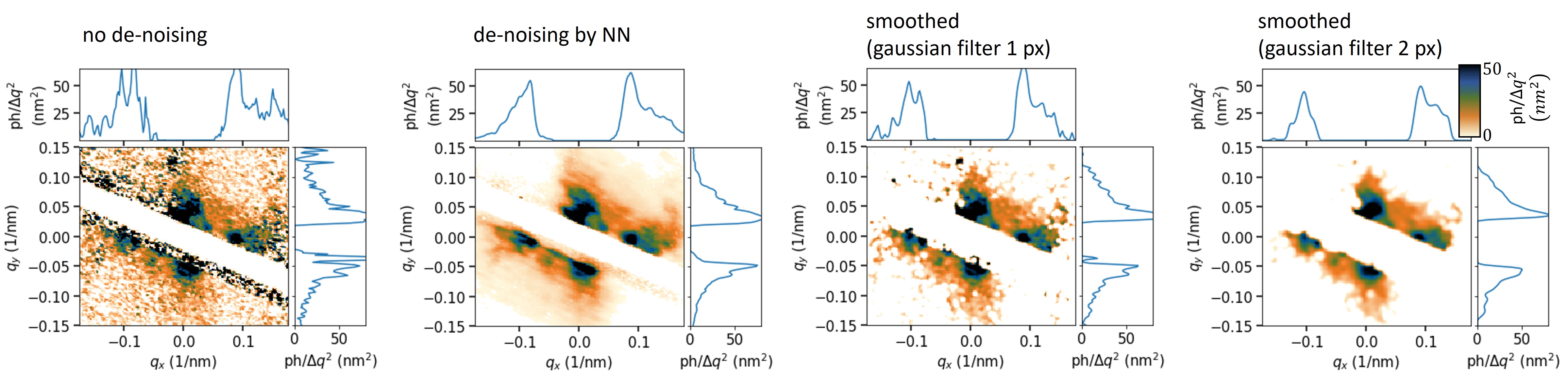}
            \small{Figure \arabic{figure}. Example for denoising of the data, exemplified on the data shown in Fig.~\ref{fig:experiment1} ($21\unit{fs}$ delay). \textbf{Left}: no denoising applied, only flat-fielding correction was used. \textbf{2nd from left}: de-noising by NN was applied. \textbf{2nd from right and right}: Same as left, but log-smoothed (i.e. the log of the data was smoothed) with a gaussian filter of width $0.002/\mathrm{nm}$ (i.e. 1 px) and  $0.004/\mathrm{nm}$ (i.e. 2 px). Note how the fine structure is preserved for de-noising using the NN compared to smoothing.}
    \end{minipage}
\end{widetext}
On this corrected data, the signal from XFEL-only pre-shot was subtracted (normalized to the main shot by gas detector measurements for the XFEL intensity), as this resembles the parasitical scattering not originating in the target: 
\begin{figure}[!h]
    \centering
    \includegraphics[width=\linewidth]{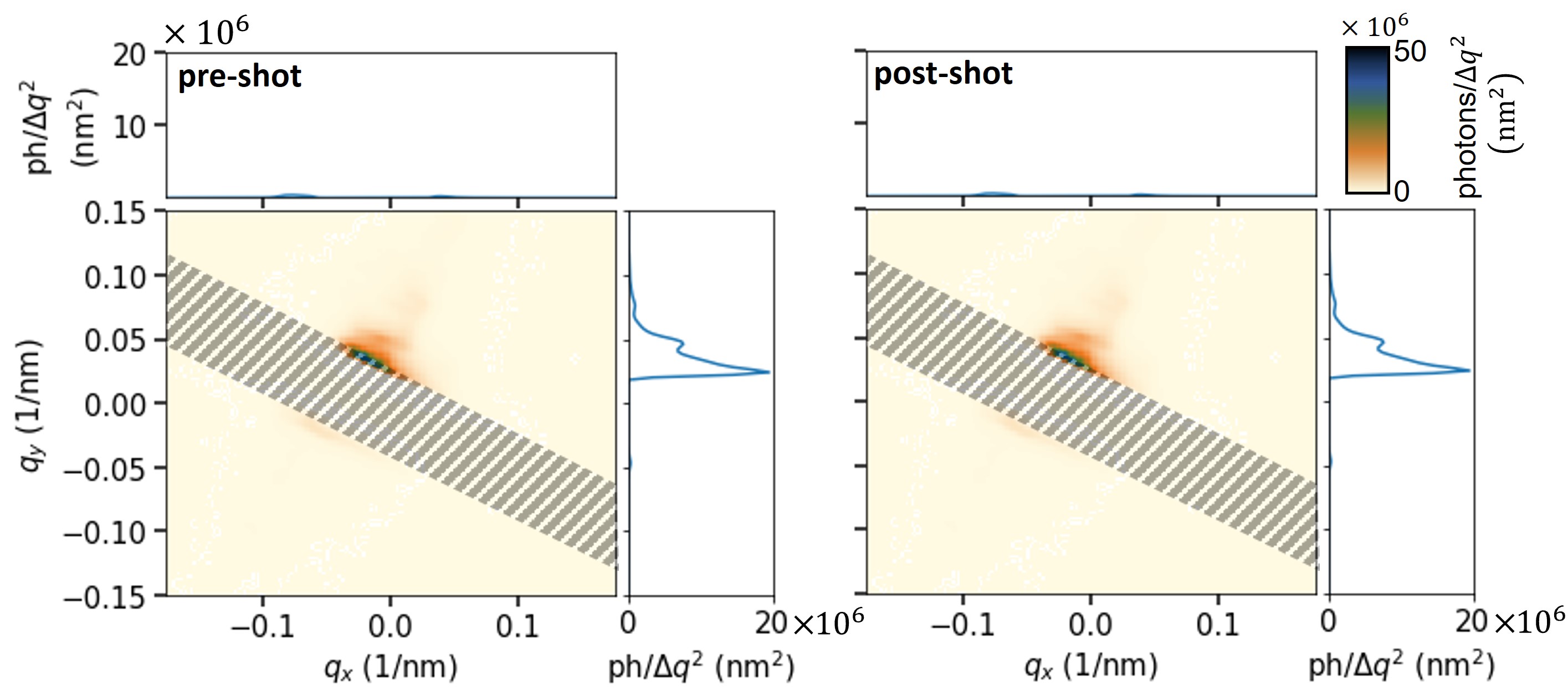}
    \caption{XFEL-only pre-shot (left) and post-shot (right) show exemplarily for the shot of delay $-20\unit{fs}$ the parasitic signal that was subtracted from the main-shots in Fig.~\ref{fig:experiment1}.}
    \label{fig:pre_postshot}
\end{figure}

\subsubsection{Quantitative analysis of scattering signal} In Fig.~\ref{fig:experiment2} the vertical/horizontal signals were projected along the perpendicular direction over a band $25\unit{px}$ wide around the peaks. Along the vertical/horizontal direction the signal was then integrated around peaks in the region where this projected signal was $2\sigma$ above the background. The errorbars in the figure indicate the background subtraction uncertainty and Poisson counting statistics. The uncertainty due to the XFEL spatial jitter is not included.

\subsection{Simulations}
\subsubsection{Simulation setup} 3D Simulations were performed using PIConGPU\cite{PIConGPU}, spatial resolution $\Delta x = \Delta y = \Delta z \approx 17 \cdot 2\pi c \omega_{pe}^{-1}$ with $\omega_{pe}=20\omega_L$, and 8 macro ions per cell. 
The silicon foil was preionized to the $+3$ state with one macro electron per macro ion (These electrons have an initial temperature $T_e=0.1 \unit{keV}$). 
Ionization was included via barrier suppression, ADK and a modified Thomas-Fermi models to correct for low temperatures, low densities as described in the PIConGPU documentation. 
A 50nm exponential preplasma was added to the front target surface to account for ASE and the laser pedestal finite contrast. Additionally, we performed a set of 2D simulations that confirmed that the qualitative results do not sensitively depend on the preplasma in the few 10s of nm scale range. The same is true for hole boring by the spatial intensity profile of the UHI laser: Previous simulations have shown no qualitative change of the dynamics in the small volume around the laser axis between simulations with and without taking into account the pulse shape\cite{Kluge2014}. 
The simulated laser is a $30\unit{fs}$ Gaussian p-polarized plane wave with $a_0=15$ peak normalized amplitude propagating at a $45^\circ$ to the target normal and is initialized $2.5 \cdot \text{FWHM}$ before and after the max intensity. 
The transversal box size $L_{\text{sim},x}=L_{\text{sim},z}=\sqrt(2)\,\lambda_L$ was chosen to match the laser phase at the periodic boundaries (in the transversal direction), $L_{\text{sim},y} = 5.08\,\lambda_L$.
The field propagation is done with the standard Yee field solver and the absorption at the boundaries in the longitudinal direction is realized with a perfectly matched layer (PML). For particles, we use a 4th order shape together with the Higuera-Cary pusher. \\

\subsubsection{Synthetic SAXS pattern} The synthetic SAXS pattern $I^{synth}$ was computed by first computing $\tilde n_e$, and taking its Fourier transform absolute square. The resulting signal was averaged over a time of $3\unit{fs}$. Note that the real XFEL pulse duration was longer, the time here was reduced due to large storage requirement of the 3D data set. 
To get a quantitative estimate of the expected photons on the detector, $I^{synth}$, the simulated X-ray signal was then scaled to the experimental intensity assuming scattering only within the ReLaX focus (FWHM) with the time average of the simulated intensity. That means, $I^{synth}$ was normalized such that $I^{synth}(\vec{Q}=\vec{0})=N_X (w_L / w_X)^2 / w_L^2 \cdot  N_e^2 \Delta \Omega$ (see Eqn.~(6) in \cite{Kluge2017}), where $N_e=410 \cdot n_c w_L^2\cdot 2\unit{\mum}$ is the estimate number of electrons within the ReLaX focus, $w_L^2$ the probed area, and $\Delta \Omega$ is the solid angle corresponding to $(\Delta q)^2=\unit{nm^{-2}}$.
Note: The similarity of the projected plasma wavelength, $\lambda_{x}^{PIC}\cong 2\pi c/\omega_{pe}/\sin{\pi/4}=60\unit{nm}$, with the vertical filamentation wavelength $\lambda_{y}^{PIC}=70\unit{nm}$ most likely is coincidence, as we repeated the simulation with a more shallow laser incidence angle of $22.5^\circ$. There, $\lambda_{x}$ is reduced as expected, while $\lambda_{y}$ remains constant. 

\subsubsection{Growth rate and filamentation wave vector} 
The instability spectrum of the fastest mode plotted in Fig.~\ref{fig:experiment1} (orange dashed line) was obtained from the PIC simulation at $t=0$. We first binned the electron momentum into bins of size $\Delta(\gamma \beta) = 0.05$. We then checked for roots of the dispersion relation Eqn.~(15) in \cite{Bret2004}, with $x$ into laser direction and $z$ into the filamentation direction. 
Eqn.~(6) in that paper was solved numerically using the Landau contour:
First, we rotated the coordinate system such that the root lies along a coordinate axis $\vec{e}_\alpha$. Then, for each $p_\alpha$ we computed and summed up the residue and integral along $p_\beta$, with symmetric bins around the root of the denominator of the second integral, again with a resolution of $\Delta(\gamma \beta) = 0.05$, and with the derivation of $f(\vec{p})$ at $\vec{p}_0$ taken symmetric around $\vec{p}_0$. 

\subsubsection{Why do the plasmons not smear out the signal in SAXS?} While in shadowgraphic methods the XFEL-propagation at a large angle w.r.t. orientation of the plasmon propagation leads to their contrast almost vanishing for probes longer than the projected plasmon period due to smearing out the density contrast, for X-ray \emph{scattering} the signal of a pattern $\tilde n(\vec{r})$ moving with velocity $v_{p,x}$ in x-direction is approximately given by the integral  over the XFEL irradiated area and XFEL pulse duration 
\begin{widetext}
    $$I(\vec{q})\propto \int_{t} \left|\int_{w}  \tilde n(\vec{r}) e^{i (q_x v_p) t} e^{-i \vec{q} \vec{r}} d\vec{r}\right|^2 dt \propto \left|FT\left(\tilde n(\vec{r})\right)\right|^2,$$
\end{widetext} i.e. simply the usual Fourier transform absolute square of the static density pattern. \\


\begin{thebibliography}{55}%
\makeatletter
\providecommand \@ifxundefined [1]{%
 \@ifx{#1\undefined}
}%
\providecommand \@ifnum [1]{%
 \ifnum #1\expandafter \@firstoftwo
 \else \expandafter \@secondoftwo
 \fi
}%
\providecommand \@ifx [1]{%
 \ifx #1\expandafter \@firstoftwo
 \else \expandafter \@secondoftwo
 \fi
}%
\providecommand \natexlab [1]{#1}%
\providecommand \enquote  [1]{``#1''}%
\providecommand \bibnamefont  [1]{#1}%
\providecommand \bibfnamefont [1]{#1}%
\providecommand \citenamefont [1]{#1}%
\providecommand \href@noop [0]{\@secondoftwo}%
\providecommand \href [0]{\begingroup \@sanitize@url \@href}%
\providecommand \@href[1]{\@@startlink{#1}\@@href}%
\providecommand \@@href[1]{\endgroup#1\@@endlink}%
\providecommand \@sanitize@url [0]{\catcode `\\12\catcode `\$12\catcode
  `\&12\catcode `\#12\catcode `\^12\catcode `\_12\catcode `\%12\relax}%
\providecommand \@@startlink[1]{}%
\providecommand \@@endlink[0]{}%
\providecommand \url  [0]{\begingroup\@sanitize@url \@url }%
\providecommand \@url [1]{\endgroup\@href {#1}{\urlprefix }}%
\providecommand \urlprefix  [0]{URL }%
\providecommand \Eprint [0]{\href }%
\providecommand \doibase [0]{http://dx.doi.org/}%
\providecommand \selectlanguage [0]{\@gobble}%
\providecommand \bibinfo  [0]{\@secondoftwo}%
\providecommand \bibfield  [0]{\@secondoftwo}%
\providecommand \translation [1]{[#1]}%
\providecommand \BibitemOpen [0]{}%
\providecommand \bibitemStop [0]{}%
\providecommand \bibitemNoStop [0]{.\EOS\space}%
\providecommand \EOS [0]{\spacefactor3000\relax}%
\providecommand \BibitemShut  [1]{\csname bibitem#1\endcsname}%
\let\auto@bib@innerbib\@empty
\bibitem [{\citenamefont {Kroll}\ \emph {et~al.}(2022)\citenamefont {Kroll},
  \citenamefont {Brack}, \citenamefont {Bernert}, \citenamefont {Bock},
  \citenamefont {Bodenstein}, \citenamefont {Br{\"{u}}chner}, \citenamefont
  {Cowan}, \citenamefont {Gaus}, \citenamefont {Gebhardt}, \citenamefont
  {Helbig}, \citenamefont {Karsch}, \citenamefont {Kluge}, \citenamefont
  {Kraft}, \citenamefont {Krause}, \citenamefont {Lessmann}, \citenamefont
  {Masood}, \citenamefont {Meister}, \citenamefont {Metzkes-Ng}, \citenamefont
  {Nossula}, \citenamefont {Pawelke}, \citenamefont {Pietzsch}, \citenamefont
  {P{\"{u}}schel}, \citenamefont {Reimold}, \citenamefont {Rehwald},
  \citenamefont {Richter}, \citenamefont {Schlenvoigt}, \citenamefont
  {Schramm}, \citenamefont {Umlandt}, \citenamefont {Ziegler}, \citenamefont
  {Zeil},\ and\ \citenamefont {Beyreuther}}]{Kroll2022}%
  \BibitemOpen
  \bibfield  {author} {\bibinfo {author} {\bibfnamefont {Florian}\ \bibnamefont
  {Kroll}}, \bibinfo {author} {\bibfnamefont {Florian-Emanuel}\ \bibnamefont
  {Brack}}, \bibinfo {author} {\bibfnamefont {Constantin}\ \bibnamefont
  {Bernert}}, \bibinfo {author} {\bibfnamefont {Stefan}\ \bibnamefont {Bock}},
  \bibinfo {author} {\bibfnamefont {Elisabeth}\ \bibnamefont {Bodenstein}},
  \bibinfo {author} {\bibfnamefont {Kerstin}\ \bibnamefont {Br{\"{u}}chner}},
  \bibinfo {author} {\bibfnamefont {Thomas~E.}\ \bibnamefont {Cowan}}, \bibinfo
  {author} {\bibfnamefont {Lennart}\ \bibnamefont {Gaus}}, \bibinfo {author}
  {\bibfnamefont {RenÃ©}\ \bibnamefont {Gebhardt}}, \bibinfo {author}
  {\bibfnamefont {Uwe}\ \bibnamefont {Helbig}}, \bibinfo {author}
  {\bibfnamefont {Leonhard}\ \bibnamefont {Karsch}}, \bibinfo {author}
  {\bibfnamefont {Thomas}\ \bibnamefont {Kluge}}, \bibinfo {author}
  {\bibfnamefont {Stephan}\ \bibnamefont {Kraft}}, \bibinfo {author}
  {\bibfnamefont {Mechthild}\ \bibnamefont {Krause}}, \bibinfo {author}
  {\bibfnamefont {Elisabeth}\ \bibnamefont {Lessmann}}, \bibinfo {author}
  {\bibfnamefont {Umar}\ \bibnamefont {Masood}}, \bibinfo {author}
  {\bibfnamefont {Sebastian}\ \bibnamefont {Meister}}, \bibinfo {author}
  {\bibfnamefont {Josefine}\ \bibnamefont {Metzkes-Ng}}, \bibinfo {author}
  {\bibfnamefont {Alexej}\ \bibnamefont {Nossula}}, \bibinfo {author}
  {\bibfnamefont {JÃ¶rg}\ \bibnamefont {Pawelke}}, \bibinfo {author}
  {\bibfnamefont {Jens}\ \bibnamefont {Pietzsch}}, \bibinfo {author}
  {\bibfnamefont {Thomas}\ \bibnamefont {P{\"{u}}schel}}, \bibinfo {author}
  {\bibfnamefont {Marvin}\ \bibnamefont {Reimold}}, \bibinfo {author}
  {\bibfnamefont {Martin}\ \bibnamefont {Rehwald}}, \bibinfo {author}
  {\bibfnamefont {Christian}\ \bibnamefont {Richter}}, \bibinfo {author}
  {\bibfnamefont {Hans-Peter}\ \bibnamefont {Schlenvoigt}}, \bibinfo {author}
  {\bibfnamefont {Ulrich}\ \bibnamefont {Schramm}}, \bibinfo {author}
  {\bibfnamefont {Marvin E.~P.}\ \bibnamefont {Umlandt}}, \bibinfo {author}
  {\bibfnamefont {Tim}\ \bibnamefont {Ziegler}}, \bibinfo {author}
  {\bibfnamefont {Karl}\ \bibnamefont {Zeil}}, \ and\ \bibinfo {author}
  {\bibfnamefont {Elke}\ \bibnamefont {Beyreuther}},\ }\bibfield  {title}
  {\enquote {\bibinfo {title} {{Tumour irradiation in mice with a
  laser-accelerated proton beam}},}\ }\href {\doibase
  10.1038/s41567-022-01520-3} {\bibfield  {journal} {\bibinfo  {journal}
  {Nature Physics}\ }\textbf {\bibinfo {volume} {18}},\ \bibinfo {pages}
  {316--322} (\bibinfo {year} {2022})}\BibitemShut {NoStop}%
\bibitem [{\citenamefont {Albert}\ \emph {et~al.}(2021)\citenamefont {Albert},
  \citenamefont {Couprie}, \citenamefont {Debus}, \citenamefont {Downer},
  \citenamefont {Faure}, \citenamefont {Flacco}, \citenamefont {Gizzi},
  \citenamefont {Grismayer}, \citenamefont {Huebl}, \citenamefont {Joshi},
  \citenamefont {Labat}, \citenamefont {Leemans}, \citenamefont {Maier},
  \citenamefont {Mangles}, \citenamefont {Mason}, \citenamefont {Mathieu},
  \citenamefont {Muggli}, \citenamefont {Nishiuchi}, \citenamefont {Osterhoff},
  \citenamefont {Rajeev}, \citenamefont {Schramm}, \citenamefont {Schreiber},
  \citenamefont {Thomas}, \citenamefont {Vay}, \citenamefont {Vranic},\ and\
  \citenamefont {Zeil}}]{Albert2021}%
  \BibitemOpen
  \bibfield  {author} {\bibinfo {author} {\bibfnamefont {FÃ©licie}\
  \bibnamefont {Albert}}, \bibinfo {author} {\bibfnamefont {M~E}\ \bibnamefont
  {Couprie}}, \bibinfo {author} {\bibfnamefont {Alexander}\ \bibnamefont
  {Debus}}, \bibinfo {author} {\bibfnamefont {Mike~C}\ \bibnamefont {Downer}},
  \bibinfo {author} {\bibfnamefont {JÃ©rÃ´me}\ \bibnamefont {Faure}}, \bibinfo
  {author} {\bibfnamefont {Alessandro}\ \bibnamefont {Flacco}}, \bibinfo
  {author} {\bibfnamefont {Leonida~A}\ \bibnamefont {Gizzi}}, \bibinfo {author}
  {\bibfnamefont {Thomas}\ \bibnamefont {Grismayer}}, \bibinfo {author}
  {\bibfnamefont {Axel}\ \bibnamefont {Huebl}}, \bibinfo {author}
  {\bibfnamefont {Chan}\ \bibnamefont {Joshi}}, \bibinfo {author}
  {\bibfnamefont {M}~\bibnamefont {Labat}}, \bibinfo {author} {\bibfnamefont
  {Wim~P}\ \bibnamefont {Leemans}}, \bibinfo {author} {\bibfnamefont
  {Andreas~R}\ \bibnamefont {Maier}}, \bibinfo {author} {\bibfnamefont {Stuart
  P~D}\ \bibnamefont {Mangles}}, \bibinfo {author} {\bibfnamefont {Paul}\
  \bibnamefont {Mason}}, \bibinfo {author} {\bibfnamefont {FranÃ§ois}\
  \bibnamefont {Mathieu}}, \bibinfo {author} {\bibfnamefont {Patric}\
  \bibnamefont {Muggli}}, \bibinfo {author} {\bibfnamefont {Mamiko}\
  \bibnamefont {Nishiuchi}}, \bibinfo {author} {\bibfnamefont {Jens}\
  \bibnamefont {Osterhoff}}, \bibinfo {author} {\bibfnamefont {P~P}\
  \bibnamefont {Rajeev}}, \bibinfo {author} {\bibfnamefont {Ulrich}\
  \bibnamefont {Schramm}}, \bibinfo {author} {\bibfnamefont {JÃ¶rg}\
  \bibnamefont {Schreiber}}, \bibinfo {author} {\bibfnamefont {Alec G~R}\
  \bibnamefont {Thomas}}, \bibinfo {author} {\bibfnamefont {Jean-Luc}\
  \bibnamefont {Vay}}, \bibinfo {author} {\bibfnamefont {Marija}\ \bibnamefont
  {Vranic}}, \ and\ \bibinfo {author} {\bibfnamefont {Karl}\ \bibnamefont
  {Zeil}},\ }\bibfield  {title} {\enquote {\bibinfo {title} {{2020 roadmap on
  plasma accelerators}},}\ }\href {\doibase 10.1088/1367-2630/abcc62}
  {\bibfield  {journal} {\bibinfo  {journal} {New Journal of Physics}\ }\textbf
  {\bibinfo {volume} {23}},\ \bibinfo {pages} {031101} (\bibinfo {year}
  {2021})}\BibitemShut {NoStop}%
\bibitem [{\citenamefont {Craxton}\ \emph {et~al.}(2015)\citenamefont
  {Craxton}, \citenamefont {Anderson}, \citenamefont {Boehly}, \citenamefont
  {Goncharov}, \citenamefont {Harding}, \citenamefont {Knauer}, \citenamefont
  {McCrory}, \citenamefont {McKenty}, \citenamefont {Meyerhofer}, \citenamefont
  {Myatt}, \citenamefont {Schmitt}, \citenamefont {Sethian}, \citenamefont
  {Short}, \citenamefont {Skupsky}, \citenamefont {Theobald}, \citenamefont
  {Kruer}, \citenamefont {Tanaka}, \citenamefont {Betti}, \citenamefont
  {Collins}, \citenamefont {Delettrez}, \citenamefont {Hu}, \citenamefont
  {Marozas}, \citenamefont {Maximov}, \citenamefont {Michel}, \citenamefont
  {Radha}, \citenamefont {Regan}, \citenamefont {Sangster}, \citenamefont
  {Seka}, \citenamefont {Solodov}, \citenamefont {Soures}, \citenamefont
  {Stoeckl},\ and\ \citenamefont {Zuegel}}]{Craxton2015}%
  \BibitemOpen
  \bibfield  {author} {\bibinfo {author} {\bibfnamefont {R.~S.}\ \bibnamefont
  {Craxton}}, \bibinfo {author} {\bibfnamefont {K.~S.}\ \bibnamefont
  {Anderson}}, \bibinfo {author} {\bibfnamefont {T.~R.}\ \bibnamefont
  {Boehly}}, \bibinfo {author} {\bibfnamefont {V.~N.}\ \bibnamefont
  {Goncharov}}, \bibinfo {author} {\bibfnamefont {D.~R.}\ \bibnamefont
  {Harding}}, \bibinfo {author} {\bibfnamefont {J.~P.}\ \bibnamefont {Knauer}},
  \bibinfo {author} {\bibfnamefont {R.~L.}\ \bibnamefont {McCrory}}, \bibinfo
  {author} {\bibfnamefont {P.~W.}\ \bibnamefont {McKenty}}, \bibinfo {author}
  {\bibfnamefont {D.~D.}\ \bibnamefont {Meyerhofer}}, \bibinfo {author}
  {\bibfnamefont {J.~F.}\ \bibnamefont {Myatt}}, \bibinfo {author}
  {\bibfnamefont {A.~J.}\ \bibnamefont {Schmitt}}, \bibinfo {author}
  {\bibfnamefont {J.~D.}\ \bibnamefont {Sethian}}, \bibinfo {author}
  {\bibfnamefont {R.~W.}\ \bibnamefont {Short}}, \bibinfo {author}
  {\bibfnamefont {S.}~\bibnamefont {Skupsky}}, \bibinfo {author} {\bibfnamefont
  {W.}~\bibnamefont {Theobald}}, \bibinfo {author} {\bibfnamefont {W.~L.}\
  \bibnamefont {Kruer}}, \bibinfo {author} {\bibfnamefont {K.}~\bibnamefont
  {Tanaka}}, \bibinfo {author} {\bibfnamefont {R.}~\bibnamefont {Betti}},
  \bibinfo {author} {\bibfnamefont {T.~J.~B.}\ \bibnamefont {Collins}},
  \bibinfo {author} {\bibfnamefont {J.~A.}\ \bibnamefont {Delettrez}}, \bibinfo
  {author} {\bibfnamefont {S.~X.}\ \bibnamefont {Hu}}, \bibinfo {author}
  {\bibfnamefont {J.~A.}\ \bibnamefont {Marozas}}, \bibinfo {author}
  {\bibfnamefont {A.~V.}\ \bibnamefont {Maximov}}, \bibinfo {author}
  {\bibfnamefont {D.~T.}\ \bibnamefont {Michel}}, \bibinfo {author}
  {\bibfnamefont {P.~B.}\ \bibnamefont {Radha}}, \bibinfo {author}
  {\bibfnamefont {S.~P.}\ \bibnamefont {Regan}}, \bibinfo {author}
  {\bibfnamefont {T.~C.}\ \bibnamefont {Sangster}}, \bibinfo {author}
  {\bibfnamefont {W.}~\bibnamefont {Seka}}, \bibinfo {author} {\bibfnamefont
  {A.~A.}\ \bibnamefont {Solodov}}, \bibinfo {author} {\bibfnamefont {J.~M.}\
  \bibnamefont {Soures}}, \bibinfo {author} {\bibfnamefont {C.}~\bibnamefont
  {Stoeckl}}, \ and\ \bibinfo {author} {\bibfnamefont {J.~D.}\ \bibnamefont
  {Zuegel}},\ }\bibfield  {title} {\enquote {\bibinfo {title} {{Direct-drive
  inertial confinement fusion: A review}},}\ }\href {\doibase
  10.1063/1.4934714} {\bibfield  {journal} {\bibinfo  {journal} {Physics of
  Plasmas}\ }\textbf {\bibinfo {volume} {22}},\ \bibinfo {pages} {110501}
  (\bibinfo {year} {2015})}\BibitemShut {NoStop}%
\bibitem [{\citenamefont {Edwards}\ and\ \citenamefont
  {Danson}(2015)}]{Edwards2015}%
  \BibitemOpen
  \bibfield  {author} {\bibinfo {author} {\bibfnamefont {C.B.}\ \bibnamefont
  {Edwards}}\ and\ \bibinfo {author} {\bibfnamefont {C.N.}\ \bibnamefont
  {Danson}},\ }\bibfield  {title} {\enquote {\bibinfo {title} {{Inertial
  confinement fusion and prospects for power production}},}\ }\href {\doibase
  10.1017/hpl.2014.51} {\bibfield  {journal} {\bibinfo  {journal} {High Power
  Laser Science and Engineering}\ }\textbf {\bibinfo {volume} {3}},\ \bibinfo
  {pages} {e4} (\bibinfo {year} {2015})}\BibitemShut {NoStop}%
\bibitem [{\citenamefont {Zylstra}\ \emph {et~al.}(2022)\citenamefont
  {Zylstra}, \citenamefont {Alexander}, \citenamefont {Bahukutumbi},
  \citenamefont {McBride}, \citenamefont {Meier}, \citenamefont {Seidl},
  \citenamefont {Wolford},\ and\ \citenamefont {Yin}}]{Zylstra2022}%
  \BibitemOpen
  \bibfield  {author} {\bibinfo {author} {\bibfnamefont {Alex}\ \bibnamefont
  {Zylstra}}, \bibinfo {author} {\bibfnamefont {Neil}\ \bibnamefont
  {Alexander}}, \bibinfo {author} {\bibfnamefont {Radha}\ \bibnamefont
  {Bahukutumbi}}, \bibinfo {author} {\bibfnamefont {Ryan}\ \bibnamefont
  {McBride}}, \bibinfo {author} {\bibfnamefont {Wayne}\ \bibnamefont {Meier}},
  \bibinfo {author} {\bibfnamefont {Peter}\ \bibnamefont {Seidl}}, \bibinfo
  {author} {\bibfnamefont {Matt}\ \bibnamefont {Wolford}}, \ and\ \bibinfo
  {author} {\bibfnamefont {Lin}\ \bibnamefont {Yin}},\ }\bibfield  {title}
  {\enquote {\bibinfo {title} {{IFE Science {\&} Technology Community Strategic
  Planning Workshop Report
  (https://lasers.llnl.gov/content/assets/docs/nif-workshops/ife-workshop-2022/IFE-Workshop-Report.pdf)}},}\
  \ }(\bibinfo {year} {2022})\BibitemShut {NoStop}%
\bibitem [{\citenamefont {Wilks}\ \emph {et~al.}(2022)\citenamefont {Wilks},
  \citenamefont {Kemp}, \citenamefont {Kirkwood}, \citenamefont {Kruer},
  \citenamefont {Ludwig}, \citenamefont {Mariscal}, \citenamefont {Marinak},
  \citenamefont {Bude}, \citenamefont {Patel}, \citenamefont {Poole},
  \citenamefont {Reagan}, \citenamefont {Rusby}, \citenamefont {Tabak},
  \citenamefont {Sherlock}, \citenamefont {Simpson}, \citenamefont {Spinka},
  \citenamefont {Tang}, \citenamefont {Williams}, \citenamefont {Friedman},
  \citenamefont {Zylstra}, \citenamefont {Ma},\ and\ \citenamefont
  {Grace}}]{Wilks2022}%
  \BibitemOpen
  \bibfield  {author} {\bibinfo {author} {\bibfnamefont {S~C}\ \bibnamefont
  {Wilks}}, \bibinfo {author} {\bibfnamefont {A~J}\ \bibnamefont {Kemp}},
  \bibinfo {author} {\bibfnamefont {R~K}\ \bibnamefont {Kirkwood}}, \bibinfo
  {author} {\bibfnamefont {W~L}\ \bibnamefont {Kruer}}, \bibinfo {author}
  {\bibfnamefont {J~D}\ \bibnamefont {Ludwig}}, \bibinfo {author}
  {\bibfnamefont {D}~\bibnamefont {Mariscal}}, \bibinfo {author} {\bibfnamefont
  {M}~\bibnamefont {Marinak}}, \bibinfo {author} {\bibfnamefont
  {J}~\bibnamefont {Bude}}, \bibinfo {author} {\bibfnamefont {P}~\bibnamefont
  {Patel}}, \bibinfo {author} {\bibfnamefont {P}~\bibnamefont {Poole}},
  \bibinfo {author} {\bibfnamefont {B}~\bibnamefont {Reagan}}, \bibinfo
  {author} {\bibfnamefont {D}~\bibnamefont {Rusby}}, \bibinfo {author}
  {\bibfnamefont {M}~\bibnamefont {Tabak}}, \bibinfo {author} {\bibfnamefont
  {M}~\bibnamefont {Sherlock}}, \bibinfo {author} {\bibfnamefont
  {R}~\bibnamefont {Simpson}}, \bibinfo {author} {\bibfnamefont
  {T}~\bibnamefont {Spinka}}, \bibinfo {author} {\bibfnamefont {V}~\bibnamefont
  {Tang}}, \bibinfo {author} {\bibfnamefont {J}~\bibnamefont {Williams}},
  \bibinfo {author} {\bibfnamefont {A}~\bibnamefont {Friedman}}, \bibinfo
  {author} {\bibfnamefont {A}~\bibnamefont {Zylstra}}, \bibinfo {author}
  {\bibfnamefont {T}~\bibnamefont {Ma}}, \ and\ \bibinfo {author}
  {\bibfnamefont {E}~\bibnamefont {Grace}},\ }\bibfield  {title} {\enquote
  {\bibinfo {title} {{Short Pulse Laser based Ion Fast Ignition for IFE
  (https://lasers.llnl.gov/content/assets/docs/nif-workshops/ife-workshop-2021/white-papers/wilks-LLNL-IFE-workshop-2022.pdf)}},}\
  }\href
  {https://lasers.llnl.gov/content/assets/docs/nif-workshops/ife-workshop-2021/white-papers/wilks-LLNL-IFE-workshop-2022.pdf}
  {\bibfield  {journal} {\bibinfo  {journal} {Whitepaper for
  LLNL-IFE-workshop}\ } (\bibinfo {year} {2022})}\BibitemShut {NoStop}%
\bibitem [{\citenamefont {Obst-Huebl}\ \emph {et~al.}(2022)\citenamefont
  {Obst-Huebl}, \citenamefont {Hakimi}, \citenamefont {Nakamura}, \citenamefont
  {Ostermayr}, \citenamefont {Tilborg}, \citenamefont {Bulanov}, \citenamefont
  {Huebl}, \citenamefont {Tang}, \citenamefont {Williams}, \citenamefont
  {Friedman}, \citenamefont {Ma},\ and\ \citenamefont
  {Grace}}]{Obst-Huebl2022}%
  \BibitemOpen
  \bibfield  {author} {\bibinfo {author} {\bibfnamefont {L.}~\bibnamefont
  {Obst-Huebl}}, \bibinfo {author} {\bibfnamefont {S}~\bibnamefont {Hakimi}},
  \bibinfo {author} {\bibfnamefont {K}~\bibnamefont {Nakamura}}, \bibinfo
  {author} {\bibfnamefont {T}~\bibnamefont {Ostermayr}}, \bibinfo {author}
  {\bibfnamefont {J~Van}\ \bibnamefont {Tilborg}}, \bibinfo {author}
  {\bibfnamefont {S}~\bibnamefont {Bulanov}}, \bibinfo {author} {\bibfnamefont
  {A}~\bibnamefont {Huebl}}, \bibinfo {author} {\bibfnamefont {V}~\bibnamefont
  {Tang}}, \bibinfo {author} {\bibfnamefont {J}~\bibnamefont {Williams}},
  \bibinfo {author} {\bibfnamefont {A}~\bibnamefont {Friedman}}, \bibinfo
  {author} {\bibfnamefont {T}~\bibnamefont {Ma}}, \ and\ \bibinfo {author}
  {\bibfnamefont {E}~\bibnamefont {Grace}},\ }\bibfield  {title} {\enquote
  {\bibinfo {title} {{BELLA PW 1 Hz Laser Experiments for Short Pulse
  Laser-based Ion Fast Ignition for IFE
  (https://lasers.llnl.gov/content/assets/docs/nif-workshops/ife-workshop-2021/white-papers/obsthuebl-LBNL-IFE-workshop-2022.pdf)}},}\
  }\href
  {https://lasers.llnl.gov/content/assets/docs/nif-workshops/ife-workshop-2021/white-papers/obsthuebl-LBNL-IFE-workshop-2022.pdf}
  {\ ,\ \bibinfo {pages} {1--10} (\bibinfo {year} {2022})}\BibitemShut
  {NoStop}%
\bibitem [{\citenamefont {Abu-Shawareb}\ \emph {et~al.}(2022)\citenamefont
  {Abu-Shawareb}, \citenamefont {Acree}, \citenamefont {Adams}, \citenamefont
  {Adams}, \citenamefont {Addis},\ and\ \citenamefont
  {Zylstra}}]{Abu-Shawareb2022}%
  \BibitemOpen
  \bibfield  {author} {\bibinfo {author} {\bibfnamefont {H.}~\bibnamefont
  {Abu-Shawareb}}, \bibinfo {author} {\bibfnamefont {R.}~\bibnamefont {Acree}},
  \bibinfo {author} {\bibfnamefont {P.}~\bibnamefont {Adams}}, \bibinfo
  {author} {\bibfnamefont {J.}~\bibnamefont {Adams}}, \bibinfo {author}
  {\bibfnamefont {B.}~\bibnamefont {Addis}}, \ and\ \bibinfo {author}
  {\bibfnamefont {A.~B.}\ \bibnamefont {Zylstra}},\ }\bibfield  {title}
  {\enquote {\bibinfo {title} {{Lawson Criterion for Ignition Exceeded in an
  Inertial Fusion Experiment}},}\ }\href {\doibase
  10.1103/PhysRevLett.129.075001} {\bibfield  {journal} {\bibinfo  {journal}
  {Physical Review Letters}\ }\textbf {\bibinfo {volume} {129}},\ \bibinfo
  {pages} {075001} (\bibinfo {year} {2022})}\BibitemShut {NoStop}%
\bibitem [{\citenamefont {Bishop (LLNL press release:
  https://www.llnl.gov/news/national-ignition-facility-achieves-fusion
  ignition)}(2022)}]{Bishop2022}%
  \BibitemOpen
  \bibfield  {author} {\bibinfo {author} {\bibfnamefont {Breanna}\ \bibnamefont
  {Bishop (LLNL press release:
  https://www.llnl.gov/news/national-ignition-facility-achieves-fusion
  ignition)}},\ }\href
  {https://www.llnl.gov/news/national-ignition-facility-achieves-fusion-ignition}
  {\enquote {\bibinfo {title} {{National Ignition Facility achieves fusion
  ignition}},}\ } (\bibinfo {year} {2022})\BibitemShut {NoStop}%
\bibitem [{\citenamefont {Glenzer}\ \emph {et~al.}(2010)\citenamefont
  {Glenzer}, \citenamefont {MacGowan}, \citenamefont {Michel}, \citenamefont
  {Meezan}, \citenamefont {Suter}, \citenamefont {Dixit}, \citenamefont
  {Kline}, \citenamefont {Kyrala}, \citenamefont {Bradley}, \citenamefont
  {Callahan}, \citenamefont {Dewald}, \citenamefont {Divol}, \citenamefont
  {Dzenitis}, \citenamefont {Edwards}, \citenamefont {Hamza}, \citenamefont
  {Haynam}, \citenamefont {Hinkel}, \citenamefont {Kalantar}, \citenamefont
  {Kilkenny}, \citenamefont {Landen}, \citenamefont {Lindl}, \citenamefont
  {LePape}, \citenamefont {Moody}, \citenamefont {Nikroo}, \citenamefont
  {Parham}, \citenamefont {Schneider}, \citenamefont {Town}, \citenamefont
  {Wegner}, \citenamefont {Widmann}, \citenamefont {Whitman}, \citenamefont
  {Young}, \citenamefont {Van~Wonterghem}, \citenamefont {Atherton},\ and\
  \citenamefont {Moses}}]{Glenzer2010}%
  \BibitemOpen
  \bibfield  {author} {\bibinfo {author} {\bibfnamefont {S.~H.}\ \bibnamefont
  {Glenzer}}, \bibinfo {author} {\bibfnamefont {B.~J.}\ \bibnamefont
  {MacGowan}}, \bibinfo {author} {\bibfnamefont {P.}~\bibnamefont {Michel}},
  \bibinfo {author} {\bibfnamefont {N.~B.}\ \bibnamefont {Meezan}}, \bibinfo
  {author} {\bibfnamefont {L.~J.}\ \bibnamefont {Suter}}, \bibinfo {author}
  {\bibfnamefont {S.~N.}\ \bibnamefont {Dixit}}, \bibinfo {author}
  {\bibfnamefont {J.~L.}\ \bibnamefont {Kline}}, \bibinfo {author}
  {\bibfnamefont {G.~A.}\ \bibnamefont {Kyrala}}, \bibinfo {author}
  {\bibfnamefont {D.~K.}\ \bibnamefont {Bradley}}, \bibinfo {author}
  {\bibfnamefont {D.~A.}\ \bibnamefont {Callahan}}, \bibinfo {author}
  {\bibfnamefont {E.~L.}\ \bibnamefont {Dewald}}, \bibinfo {author}
  {\bibfnamefont {L.}~\bibnamefont {Divol}}, \bibinfo {author} {\bibfnamefont
  {E.}~\bibnamefont {Dzenitis}}, \bibinfo {author} {\bibfnamefont {M.~J.}\
  \bibnamefont {Edwards}}, \bibinfo {author} {\bibfnamefont {A.~V.}\
  \bibnamefont {Hamza}}, \bibinfo {author} {\bibfnamefont {C.~A.}\ \bibnamefont
  {Haynam}}, \bibinfo {author} {\bibfnamefont {D.~E.}\ \bibnamefont {Hinkel}},
  \bibinfo {author} {\bibfnamefont {D.~H.}\ \bibnamefont {Kalantar}}, \bibinfo
  {author} {\bibfnamefont {J.~D.}\ \bibnamefont {Kilkenny}}, \bibinfo {author}
  {\bibfnamefont {O.~L.}\ \bibnamefont {Landen}}, \bibinfo {author}
  {\bibfnamefont {J.~D.}\ \bibnamefont {Lindl}}, \bibinfo {author}
  {\bibfnamefont {S.}~\bibnamefont {LePape}}, \bibinfo {author} {\bibfnamefont
  {J.~D.}\ \bibnamefont {Moody}}, \bibinfo {author} {\bibfnamefont
  {A.}~\bibnamefont {Nikroo}}, \bibinfo {author} {\bibfnamefont
  {T.}~\bibnamefont {Parham}}, \bibinfo {author} {\bibfnamefont {M.~B.}\
  \bibnamefont {Schneider}}, \bibinfo {author} {\bibfnamefont {R.~P.~J.}\
  \bibnamefont {Town}}, \bibinfo {author} {\bibfnamefont {P.}~\bibnamefont
  {Wegner}}, \bibinfo {author} {\bibfnamefont {K.}~\bibnamefont {Widmann}},
  \bibinfo {author} {\bibfnamefont {P.}~\bibnamefont {Whitman}}, \bibinfo
  {author} {\bibfnamefont {B.~K.~F.}\ \bibnamefont {Young}}, \bibinfo {author}
  {\bibfnamefont {B.}~\bibnamefont {Van~Wonterghem}}, \bibinfo {author}
  {\bibfnamefont {L.~J.}\ \bibnamefont {Atherton}}, \ and\ \bibinfo {author}
  {\bibfnamefont {E.~I.}\ \bibnamefont {Moses}},\ }\bibfield  {title} {\enquote
  {\bibinfo {title} {{Symmetric Inertial Confinement Fusion Implosions at
  Ultra-High Laser Energies}},}\ }\href {\doibase 10.1126/science.1185634}
  {\bibfield  {journal} {\bibinfo  {journal} {Science}\ }\textbf {\bibinfo
  {volume} {327}},\ \bibinfo {pages} {1228--1231} (\bibinfo {year}
  {2010})}\BibitemShut {NoStop}%
\bibitem [{\citenamefont {Tabak}\ \emph {et~al.}(1994)\citenamefont {Tabak},
  \citenamefont {Hammer}, \citenamefont {Glinsky}, \citenamefont {Kruer},
  \citenamefont {Wilks}, \citenamefont {Woodworth}, \citenamefont {Campbell},
  \citenamefont {Perry},\ and\ \citenamefont {Mason}}]{Tabak1994}%
  \BibitemOpen
  \bibfield  {author} {\bibinfo {author} {\bibfnamefont {Max}\ \bibnamefont
  {Tabak}}, \bibinfo {author} {\bibfnamefont {James}\ \bibnamefont {Hammer}},
  \bibinfo {author} {\bibfnamefont {Michael~E.}\ \bibnamefont {Glinsky}},
  \bibinfo {author} {\bibfnamefont {William~L.}\ \bibnamefont {Kruer}},
  \bibinfo {author} {\bibfnamefont {Scott~C.}\ \bibnamefont {Wilks}}, \bibinfo
  {author} {\bibfnamefont {John}\ \bibnamefont {Woodworth}}, \bibinfo {author}
  {\bibfnamefont {E.~Michael}\ \bibnamefont {Campbell}}, \bibinfo {author}
  {\bibfnamefont {Michael~D.}\ \bibnamefont {Perry}}, \ and\ \bibinfo {author}
  {\bibfnamefont {Rodney~J.}\ \bibnamefont {Mason}},\ }\bibfield  {title}
  {\enquote {\bibinfo {title} {{Ignition and high gain with ultrapowerful
  lasers}},}\ }\href {\doibase 10.1063/1.870664} {\bibfield  {journal}
  {\bibinfo  {journal} {Physics of Plasmas}\ }\textbf {\bibinfo {volume} {1}},\
  \bibinfo {pages} {1626--1634} (\bibinfo {year} {1994})}\BibitemShut {NoStop}%
\bibitem [{\citenamefont {Roth}\ \emph {et~al.}(2001)\citenamefont {Roth},
  \citenamefont {Cowan}, \citenamefont {Key}, \citenamefont {Hatchett},
  \citenamefont {Brown}, \citenamefont {Fountain}, \citenamefont {Johnson},
  \citenamefont {Pennington}, \citenamefont {Snavely}, \citenamefont {Wilks},
  \citenamefont {Yasuike}, \citenamefont {Ruhl}, \citenamefont {Pegoraro},
  \citenamefont {Bulanov}, \citenamefont {Campbell}, \citenamefont {Perry},\
  and\ \citenamefont {Powell}}]{Roth2001}%
  \BibitemOpen
  \bibfield  {author} {\bibinfo {author} {\bibfnamefont {M.}~\bibnamefont
  {Roth}}, \bibinfo {author} {\bibfnamefont {T.~E.}\ \bibnamefont {Cowan}},
  \bibinfo {author} {\bibfnamefont {M.~H.}\ \bibnamefont {Key}}, \bibinfo
  {author} {\bibfnamefont {S.~P.}\ \bibnamefont {Hatchett}}, \bibinfo {author}
  {\bibfnamefont {C.}~\bibnamefont {Brown}}, \bibinfo {author} {\bibfnamefont
  {W.}~\bibnamefont {Fountain}}, \bibinfo {author} {\bibfnamefont
  {J.}~\bibnamefont {Johnson}}, \bibinfo {author} {\bibfnamefont {D.~M.}\
  \bibnamefont {Pennington}}, \bibinfo {author} {\bibfnamefont {R.~A.}\
  \bibnamefont {Snavely}}, \bibinfo {author} {\bibfnamefont {S.~C.}\
  \bibnamefont {Wilks}}, \bibinfo {author} {\bibfnamefont {K.}~\bibnamefont
  {Yasuike}}, \bibinfo {author} {\bibfnamefont {H.}~\bibnamefont {Ruhl}},
  \bibinfo {author} {\bibfnamefont {F.}~\bibnamefont {Pegoraro}}, \bibinfo
  {author} {\bibfnamefont {S.~V.}\ \bibnamefont {Bulanov}}, \bibinfo {author}
  {\bibfnamefont {E.~M.}\ \bibnamefont {Campbell}}, \bibinfo {author}
  {\bibfnamefont {M.~D.}\ \bibnamefont {Perry}}, \ and\ \bibinfo {author}
  {\bibfnamefont {H.}~\bibnamefont {Powell}},\ }\bibfield  {title} {\enquote
  {\bibinfo {title} {{Fast ignition by intense laser-accelerated proton
  beams}},}\ }\href {\doibase 10.1103/PhysRevLett.86.436} {\bibfield  {journal}
  {\bibinfo  {journal} {Physical Review Letters}\ }\textbf {\bibinfo {volume}
  {86}},\ \bibinfo {pages} {436--439} (\bibinfo {year} {2001})}\BibitemShut
  {NoStop}%
\bibitem [{\citenamefont {Campbell}\ \emph {et~al.}(2021)\citenamefont
  {Campbell}, \citenamefont {Sangster}, \citenamefont {Goncharov},
  \citenamefont {Zuegel}, \citenamefont {Morse}, \citenamefont {Sorce},
  \citenamefont {Collins}, \citenamefont {Wei}, \citenamefont {Betti},
  \citenamefont {Regan}, \citenamefont {Froula}, \citenamefont {Dorrer},
  \citenamefont {Harding}, \citenamefont {Gopalaswamy}, \citenamefont {Knauer},
  \citenamefont {Shah}, \citenamefont {Mannion}, \citenamefont {Marozas},
  \citenamefont {Radha}, \citenamefont {Rosenberg}, \citenamefont {Collins},
  \citenamefont {Christopherson}, \citenamefont {Solodov}, \citenamefont {Cao},
  \citenamefont {Palastro}, \citenamefont {Follett},\ and\ \citenamefont
  {Farrell}}]{Campbell2021}%
  \BibitemOpen
  \bibfield  {author} {\bibinfo {author} {\bibfnamefont {E.~M.}\ \bibnamefont
  {Campbell}}, \bibinfo {author} {\bibfnamefont {T.~C.}\ \bibnamefont
  {Sangster}}, \bibinfo {author} {\bibfnamefont {V.~N.}\ \bibnamefont
  {Goncharov}}, \bibinfo {author} {\bibfnamefont {J.~D.}\ \bibnamefont
  {Zuegel}}, \bibinfo {author} {\bibfnamefont {S.~F.B.}\ \bibnamefont {Morse}},
  \bibinfo {author} {\bibfnamefont {C.}~\bibnamefont {Sorce}}, \bibinfo
  {author} {\bibfnamefont {G.~W.}\ \bibnamefont {Collins}}, \bibinfo {author}
  {\bibfnamefont {M.~S.}\ \bibnamefont {Wei}}, \bibinfo {author} {\bibfnamefont
  {R.}~\bibnamefont {Betti}}, \bibinfo {author} {\bibfnamefont {S.~P.}\
  \bibnamefont {Regan}}, \bibinfo {author} {\bibfnamefont {D.~H.}\ \bibnamefont
  {Froula}}, \bibinfo {author} {\bibfnamefont {C.}~\bibnamefont {Dorrer}},
  \bibinfo {author} {\bibfnamefont {D.~R.}\ \bibnamefont {Harding}}, \bibinfo
  {author} {\bibfnamefont {V.}~\bibnamefont {Gopalaswamy}}, \bibinfo {author}
  {\bibfnamefont {J.~P.}\ \bibnamefont {Knauer}}, \bibinfo {author}
  {\bibfnamefont {R.}~\bibnamefont {Shah}}, \bibinfo {author} {\bibfnamefont
  {O.~M.}\ \bibnamefont {Mannion}}, \bibinfo {author} {\bibfnamefont {J.~A.}\
  \bibnamefont {Marozas}}, \bibinfo {author} {\bibfnamefont {P.~B.}\
  \bibnamefont {Radha}}, \bibinfo {author} {\bibfnamefont {M.~J.}\ \bibnamefont
  {Rosenberg}}, \bibinfo {author} {\bibfnamefont {T.~J.B.}\ \bibnamefont
  {Collins}}, \bibinfo {author} {\bibfnamefont {A.~R.}\ \bibnamefont
  {Christopherson}}, \bibinfo {author} {\bibfnamefont {A.~A.}\ \bibnamefont
  {Solodov}}, \bibinfo {author} {\bibfnamefont {D.}~\bibnamefont {Cao}},
  \bibinfo {author} {\bibfnamefont {J.~P.}\ \bibnamefont {Palastro}}, \bibinfo
  {author} {\bibfnamefont {R.~K.}\ \bibnamefont {Follett}}, \ and\ \bibinfo
  {author} {\bibfnamefont {M.}~\bibnamefont {Farrell}},\ }\bibfield  {title}
  {\enquote {\bibinfo {title} {{Direct-drive laser fusion: status, plans and
  future}},}\ }\href {\doibase 10.1098/RSTA.2020.0011} {\bibfield  {journal}
  {\bibinfo  {journal} {Philosophical Transactions of the Royal Society A}\
  }\textbf {\bibinfo {volume} {379}} (\bibinfo {year} {2021}),\
  10.1098/RSTA.2020.0011}\BibitemShut {NoStop}%
\bibitem [{\citenamefont {Mahdavi}\ and\ \citenamefont
  {Khanzadeh}(2021)}]{Mahdavi2021}%
  \BibitemOpen
  \bibfield  {author} {\bibinfo {author} {\bibfnamefont {M.}~\bibnamefont
  {Mahdavi}}\ and\ \bibinfo {author} {\bibfnamefont {H.}~\bibnamefont
  {Khanzadeh}},\ }\bibfield  {title} {\enquote {\bibinfo {title} {{The effect
  of the plasma density gradient on current filamentation instability}},}\
  }\href {\doibase 10.1002/ctpp.202100027} {\bibfield  {journal} {\bibinfo
  {journal} {Contributions to Plasma Physics}\ }\textbf {\bibinfo {volume}
  {61}} (\bibinfo {year} {2021}),\ 10.1002/ctpp.202100027}\BibitemShut
  {NoStop}%
\bibitem [{\citenamefont {Badziak}\ and\ \citenamefont
  {Doma{\'{n}}ski}(2023)}]{Badziak2023}%
  \BibitemOpen
  \bibfield  {author} {\bibinfo {author} {\bibfnamefont {J.}~\bibnamefont
  {Badziak}}\ and\ \bibinfo {author} {\bibfnamefont {J.}~\bibnamefont
  {Doma{\'{n}}ski}},\ }\bibfield  {title} {\enquote {\bibinfo {title} {{In
  search of ways to improve the properties of a laser-accelerated heavy ion
  beam relevant for fusion fast ignition}},}\ }\href {\doibase
  10.1063/5.0147298} {\bibfield  {journal} {\bibinfo  {journal} {Physics of
  Plasmas}\ }\textbf {\bibinfo {volume} {30}} (\bibinfo {year} {2023}),\
  10.1063/5.0147298}\BibitemShut {NoStop}%
\bibitem [{\citenamefont {Bret}\ \emph {et~al.}(2004)\citenamefont {Bret},
  \citenamefont {Firpo},\ and\ \citenamefont {Deutsch}}]{Bret2004}%
  \BibitemOpen
  \bibfield  {author} {\bibinfo {author} {\bibfnamefont {A.}~\bibnamefont
  {Bret}}, \bibinfo {author} {\bibfnamefont {M.~C.}\ \bibnamefont {Firpo}}, \
  and\ \bibinfo {author} {\bibfnamefont {C.}~\bibnamefont {Deutsch}},\
  }\bibfield  {title} {\enquote {\bibinfo {title} {{Collective electromagnetic
  modes for beam-plasma interaction in the whole [Formula presented] space}},}\
  }\href {\doibase 10.1103/PhysRevE.70.046401} {\bibfield  {journal} {\bibinfo
  {journal} {Physical Review E - Statistical Physics, Plasmas, Fluids, and
  Related Interdisciplinary Topics}\ }\textbf {\bibinfo {volume} {70}},\
  \bibinfo {pages} {15} (\bibinfo {year} {2004})}\BibitemShut {NoStop}%
\bibitem [{\citenamefont {Metzkes}\ \emph {et~al.}(2014)\citenamefont
  {Metzkes}, \citenamefont {Kluge}, \citenamefont {Zeil}, \citenamefont
  {Bussmann}, \citenamefont {Kraft}, \citenamefont {Cowan},\ and\ \citenamefont
  {Schramm}}]{Metzkes2014a}%
  \BibitemOpen
  \bibfield  {author} {\bibinfo {author} {\bibfnamefont {J.}~\bibnamefont
  {Metzkes}}, \bibinfo {author} {\bibfnamefont {T.}~\bibnamefont {Kluge}},
  \bibinfo {author} {\bibfnamefont {K.}~\bibnamefont {Zeil}}, \bibinfo {author}
  {\bibfnamefont {M.}~\bibnamefont {Bussmann}}, \bibinfo {author}
  {\bibfnamefont {S.~D.}\ \bibnamefont {Kraft}}, \bibinfo {author}
  {\bibfnamefont {T.~E.}\ \bibnamefont {Cowan}}, \ and\ \bibinfo {author}
  {\bibfnamefont {U.}~\bibnamefont {Schramm}},\ }\bibfield  {title} {\enquote
  {\bibinfo {title} {{Experimental observation of transverse modulations in
  laser-driven proton beams}},}\ }\href {\doibase
  10.1088/1367-2630/16/2/023008} {\bibfield  {journal} {\bibinfo  {journal}
  {New Journal of Physics}\ }\textbf {\bibinfo {volume} {16}},\ \bibinfo
  {pages} {23008} (\bibinfo {year} {2014})}\BibitemShut {NoStop}%
\bibitem [{\citenamefont {G{\"{o}}de}\ \emph {et~al.}(2017)\citenamefont
  {G{\"{o}}de}, \citenamefont {R{\"{o}}del}, \citenamefont {Zeil},
  \citenamefont {Mishra}, \citenamefont {Gauthier}, \citenamefont {Brack},
  \citenamefont {Kluge}, \citenamefont {MacDonald}, \citenamefont {Metzkes},
  \citenamefont {Obst}, \citenamefont {Rehwald}, \citenamefont {Ruyer},
  \citenamefont {Schlenvoigt}, \citenamefont {Schumaker}, \citenamefont
  {Sommer}, \citenamefont {Cowan}, \citenamefont {Schramm}, \citenamefont
  {Glenzer},\ and\ \citenamefont {Fiuza}}]{Gode2017}%
  \BibitemOpen
  \bibfield  {author} {\bibinfo {author} {\bibfnamefont {S.}~\bibnamefont
  {G{\"{o}}de}}, \bibinfo {author} {\bibfnamefont {C.}~\bibnamefont
  {R{\"{o}}del}}, \bibinfo {author} {\bibfnamefont {K.}~\bibnamefont {Zeil}},
  \bibinfo {author} {\bibfnamefont {R.}~\bibnamefont {Mishra}}, \bibinfo
  {author} {\bibfnamefont {M.}~\bibnamefont {Gauthier}}, \bibinfo {author}
  {\bibfnamefont {F.-E.}\ \bibnamefont {Brack}}, \bibinfo {author}
  {\bibfnamefont {T.}~\bibnamefont {Kluge}}, \bibinfo {author} {\bibfnamefont
  {M.J.~J.}\ \bibnamefont {MacDonald}}, \bibinfo {author} {\bibfnamefont
  {J.}~\bibnamefont {Metzkes}}, \bibinfo {author} {\bibfnamefont
  {L.}~\bibnamefont {Obst}}, \bibinfo {author} {\bibfnamefont {M.}~\bibnamefont
  {Rehwald}}, \bibinfo {author} {\bibfnamefont {C.}~\bibnamefont {Ruyer}},
  \bibinfo {author} {\bibfnamefont {H.-P.~P.}\ \bibnamefont {Schlenvoigt}},
  \bibinfo {author} {\bibfnamefont {W.}~\bibnamefont {Schumaker}}, \bibinfo
  {author} {\bibfnamefont {P.}~\bibnamefont {Sommer}}, \bibinfo {author}
  {\bibfnamefont {T.E.~E.}\ \bibnamefont {Cowan}}, \bibinfo {author}
  {\bibfnamefont {U.}~\bibnamefont {Schramm}}, \bibinfo {author} {\bibfnamefont
  {S.}~\bibnamefont {Glenzer}}, \ and\ \bibinfo {author} {\bibfnamefont
  {F.}~\bibnamefont {Fiuza}},\ }\bibfield  {title} {\enquote {\bibinfo {title}
  {{Relativistic electron streaming instabilities modulate proton beams
  accelerated in laser-plasma interactions}},}\ }\href {\doibase
  10.1103/PhysRevLett.118.194801} {\bibfield  {journal} {\bibinfo  {journal}
  {Physical Review Letters}\ }\textbf {\bibinfo {volume} {118}},\ \bibinfo
  {pages} {194801} (\bibinfo {year} {2017})}\BibitemShut {NoStop}%
\bibitem [{\citenamefont {Scott}\ \emph {et~al.}(2017)\citenamefont {Scott},
  \citenamefont {Brenner}, \citenamefont {Bagnoud}, \citenamefont {Clarke},
  \citenamefont {Gonzalez-Izquierdo}, \citenamefont {Green}, \citenamefont
  {Heathcote}, \citenamefont {Powell}, \citenamefont {Rusby}, \citenamefont
  {Zielbauer}, \citenamefont {McKenna},\ and\ \citenamefont
  {Neely}}]{Scott2017}%
  \BibitemOpen
  \bibfield  {author} {\bibinfo {author} {\bibfnamefont {G.~G.}\ \bibnamefont
  {Scott}}, \bibinfo {author} {\bibfnamefont {C.~M.}\ \bibnamefont {Brenner}},
  \bibinfo {author} {\bibfnamefont {V.}~\bibnamefont {Bagnoud}}, \bibinfo
  {author} {\bibfnamefont {R.~J.}\ \bibnamefont {Clarke}}, \bibinfo {author}
  {\bibfnamefont {B.}~\bibnamefont {Gonzalez-Izquierdo}}, \bibinfo {author}
  {\bibfnamefont {J.~S.}\ \bibnamefont {Green}}, \bibinfo {author}
  {\bibfnamefont {R.~I.}\ \bibnamefont {Heathcote}}, \bibinfo {author}
  {\bibfnamefont {H.~W.}\ \bibnamefont {Powell}}, \bibinfo {author}
  {\bibfnamefont {D.~R.}\ \bibnamefont {Rusby}}, \bibinfo {author}
  {\bibfnamefont {B.}~\bibnamefont {Zielbauer}}, \bibinfo {author}
  {\bibfnamefont {P.}~\bibnamefont {McKenna}}, \ and\ \bibinfo {author}
  {\bibfnamefont {D.}~\bibnamefont {Neely}},\ }\bibfield  {title} {\enquote
  {\bibinfo {title} {{Diagnosis of Weibel instability evolution in the rear
  surface density scale lengths of laser solid interactions via proton
  acceleration}},}\ }\href {\doibase 10.1088/1367-2630/aa652c} {\bibfield
  {journal} {\bibinfo  {journal} {New Journal of Physics}\ }\textbf {\bibinfo
  {volume} {19}},\ \bibinfo {pages} {043010} (\bibinfo {year}
  {2017})}\BibitemShut {NoStop}%
\bibitem [{\citenamefont {Huang}\ \emph {et~al.}(2017)\citenamefont {Huang},
  \citenamefont {Schlenvoigt}, \citenamefont {Takabe},\ and\ \citenamefont
  {Cowan}}]{Huang2017}%
  \BibitemOpen
  \bibfield  {author} {\bibinfo {author} {\bibfnamefont {L.~G.}\ \bibnamefont
  {Huang}}, \bibinfo {author} {\bibfnamefont {H.~P.}\ \bibnamefont
  {Schlenvoigt}}, \bibinfo {author} {\bibfnamefont {H.}~\bibnamefont {Takabe}},
  \ and\ \bibinfo {author} {\bibfnamefont {T.~E.}\ \bibnamefont {Cowan}},\
  }\bibfield  {title} {\enquote {\bibinfo {title} {{Ionization and reflux
  dependence of magnetic instability generation and probing inside
  laser-irradiated solid thin foils}},}\ }\href {\doibase 10.1063/1.4989457}
  {\bibfield  {journal} {\bibinfo  {journal} {Physics of Plasmas}\ }\textbf
  {\bibinfo {volume} {24}},\ \bibinfo {pages} {103115} (\bibinfo {year}
  {2017})}\BibitemShut {NoStop}%
\bibitem [{\citenamefont {Macchi}\ \emph {et~al.}(2002)\citenamefont {Macchi},
  \citenamefont {Cornolti},\ and\ \citenamefont {Pegoraro}}]{Macchi2002}%
  \BibitemOpen
  \bibfield  {author} {\bibinfo {author} {\bibfnamefont {Andrea}\ \bibnamefont
  {Macchi}}, \bibinfo {author} {\bibfnamefont {Fulvio}\ \bibnamefont
  {Cornolti}}, \ and\ \bibinfo {author} {\bibfnamefont {Francesco}\
  \bibnamefont {Pegoraro}},\ }\bibfield  {title} {\enquote {\bibinfo {title}
  {{Two-surface wave decay}},}\ }\href {\doibase 10.1063/1.1464146} {\bibfield
  {journal} {\bibinfo  {journal} {Physics of Plasmas}\ }\textbf {\bibinfo
  {volume} {9}},\ \bibinfo {pages} {1704--1711} (\bibinfo {year}
  {2002})}\BibitemShut {NoStop}%
\bibitem [{\citenamefont {Palmer}\ \emph {et~al.}(2012)\citenamefont {Palmer},
  \citenamefont {Schreiber}, \citenamefont {Nagel}, \citenamefont {Dover},
  \citenamefont {Bellei}, \citenamefont {Beg}, \citenamefont {Bott},
  \citenamefont {Clarke}, \citenamefont {Dangor}, \citenamefont {Hassan},
  \citenamefont {Hilz}, \citenamefont {Jung}, \citenamefont {Kneip},
  \citenamefont {Mangles}, \citenamefont {Lancaster}, \citenamefont {Rehman},
  \citenamefont {Robinson}, \citenamefont {Spindloe}, \citenamefont {Szerypo},
  \citenamefont {Tatarakis}, \citenamefont {Yeung}, \citenamefont {Zepf},\ and\
  \citenamefont {Najmudin}}]{Palmer2012}%
  \BibitemOpen
  \bibfield  {author} {\bibinfo {author} {\bibfnamefont {C.~A.J.}\ \bibnamefont
  {Palmer}}, \bibinfo {author} {\bibfnamefont {J.}~\bibnamefont {Schreiber}},
  \bibinfo {author} {\bibfnamefont {S.~R.}\ \bibnamefont {Nagel}}, \bibinfo
  {author} {\bibfnamefont {N.~P.}\ \bibnamefont {Dover}}, \bibinfo {author}
  {\bibfnamefont {C.}~\bibnamefont {Bellei}}, \bibinfo {author} {\bibfnamefont
  {F.~N.}\ \bibnamefont {Beg}}, \bibinfo {author} {\bibfnamefont
  {S.}~\bibnamefont {Bott}}, \bibinfo {author} {\bibfnamefont {R.~J.}\
  \bibnamefont {Clarke}}, \bibinfo {author} {\bibfnamefont {A.~E.}\
  \bibnamefont {Dangor}}, \bibinfo {author} {\bibfnamefont {S.~M.}\
  \bibnamefont {Hassan}}, \bibinfo {author} {\bibfnamefont {P.}~\bibnamefont
  {Hilz}}, \bibinfo {author} {\bibfnamefont {D.}~\bibnamefont {Jung}}, \bibinfo
  {author} {\bibfnamefont {S.}~\bibnamefont {Kneip}}, \bibinfo {author}
  {\bibfnamefont {S.~P.D.}\ \bibnamefont {Mangles}}, \bibinfo {author}
  {\bibfnamefont {K.~L.}\ \bibnamefont {Lancaster}}, \bibinfo {author}
  {\bibfnamefont {A.}~\bibnamefont {Rehman}}, \bibinfo {author} {\bibfnamefont
  {A.~P.L.}\ \bibnamefont {Robinson}}, \bibinfo {author} {\bibfnamefont
  {C.}~\bibnamefont {Spindloe}}, \bibinfo {author} {\bibfnamefont
  {J.}~\bibnamefont {Szerypo}}, \bibinfo {author} {\bibfnamefont
  {M.}~\bibnamefont {Tatarakis}}, \bibinfo {author} {\bibfnamefont
  {M.}~\bibnamefont {Yeung}}, \bibinfo {author} {\bibfnamefont
  {M.}~\bibnamefont {Zepf}}, \ and\ \bibinfo {author} {\bibfnamefont
  {Z.}~\bibnamefont {Najmudin}},\ }\bibfield  {title} {\enquote {\bibinfo
  {title} {{Rayleigh-Taylor instability of an ultrathin foil accelerated by the
  radiation pressure of an intense laser}},}\ }\href {\doibase
  10.1103/PhysRevLett.108.225002} {\bibfield  {journal} {\bibinfo  {journal}
  {Physical Review Letters}\ }\textbf {\bibinfo {volume} {108}},\ \bibinfo
  {pages} {225002} (\bibinfo {year} {2012})}\BibitemShut {NoStop}%
\bibitem [{\citenamefont {Kluge}\ \emph {et~al.}(2015)\citenamefont {Kluge},
  \citenamefont {Metzkes}, \citenamefont {Zeil}, \citenamefont {Bussmann},
  \citenamefont {Schramm},\ and\ \citenamefont {Cowan}}]{Kluge2015}%
  \BibitemOpen
  \bibfield  {author} {\bibinfo {author} {\bibfnamefont {T.}~\bibnamefont
  {Kluge}}, \bibinfo {author} {\bibfnamefont {J.}~\bibnamefont {Metzkes}},
  \bibinfo {author} {\bibfnamefont {K.}~\bibnamefont {Zeil}}, \bibinfo {author}
  {\bibfnamefont {M.}~\bibnamefont {Bussmann}}, \bibinfo {author}
  {\bibfnamefont {U.}~\bibnamefont {Schramm}}, \ and\ \bibinfo {author}
  {\bibfnamefont {T.~E.}\ \bibnamefont {Cowan}},\ }\bibfield  {title} {\enquote
  {\bibinfo {title} {{Two surface plasmon decay of plasma oscillations}},}\
  }\href {\doibase 10.1063/1.4922673} {\bibfield  {journal} {\bibinfo
  {journal} {Physics of Plasmas}\ }\textbf {\bibinfo {volume} {22}},\ \bibinfo
  {pages} {64502} (\bibinfo {year} {2015})}\BibitemShut {NoStop}%
\bibitem [{\citenamefont {Sgattoni}\ \emph {et~al.}(2015)\citenamefont
  {Sgattoni}, \citenamefont {Sinigardi}, \citenamefont {Fedeli}, \citenamefont
  {Pegoraro},\ and\ \citenamefont {Macchi}}]{Sgattoni2015}%
  \BibitemOpen
  \bibfield  {author} {\bibinfo {author} {\bibfnamefont {A.}~\bibnamefont
  {Sgattoni}}, \bibinfo {author} {\bibfnamefont {S.}~\bibnamefont {Sinigardi}},
  \bibinfo {author} {\bibfnamefont {L.}~\bibnamefont {Fedeli}}, \bibinfo
  {author} {\bibfnamefont {F.}~\bibnamefont {Pegoraro}}, \ and\ \bibinfo
  {author} {\bibfnamefont {A.}~\bibnamefont {Macchi}},\ }\bibfield  {title}
  {\enquote {\bibinfo {title} {{Laser-driven Rayleigh-Taylor instability:
  Plasmonic effects and three-dimensional structures}},}\ }\href {\doibase
  10.1103/PhysRevE.91.013106} {\bibfield  {journal} {\bibinfo  {journal}
  {Physical Review E - Statistical, Nonlinear, and Soft Matter Physics}\
  }\textbf {\bibinfo {volume} {91}},\ \bibinfo {pages} {013106} (\bibinfo
  {year} {2015})}\BibitemShut {NoStop}%
\bibitem [{\citenamefont {Sokolowski-Tinten}\ \emph {et~al.}(1998)\citenamefont
  {Sokolowski-Tinten}, \citenamefont {Bialkowski}, \citenamefont {Cavalleri},
  \citenamefont {Von~der Linde}, \citenamefont {Oparin}, \citenamefont
  {Meyer-Ter-Vehn},\ and\ \citenamefont {Anisimov}}]{Sokolowski-Tinten1998}%
  \BibitemOpen
  \bibfield  {author} {\bibinfo {author} {\bibfnamefont {K.}~\bibnamefont
  {Sokolowski-Tinten}}, \bibinfo {author} {\bibfnamefont {J.}~\bibnamefont
  {Bialkowski}}, \bibinfo {author} {\bibfnamefont {A.}~\bibnamefont
  {Cavalleri}}, \bibinfo {author} {\bibfnamefont {D.}~\bibnamefont {Von~der
  Linde}}, \bibinfo {author} {\bibfnamefont {A.}~\bibnamefont {Oparin}},
  \bibinfo {author} {\bibfnamefont {J.}~\bibnamefont {Meyer-Ter-Vehn}}, \ and\
  \bibinfo {author} {\bibfnamefont {S.~I.}\ \bibnamefont {Anisimov}},\
  }\bibfield  {title} {\enquote {\bibinfo {title} {{Transient States of Matter
  during Short Pulse Laser Ablation}},}\ }\href {\doibase
  10.1103/PhysRevLett.81.224} {\bibfield  {journal} {\bibinfo  {journal}
  {Physical Review Letters}\ }\textbf {\bibinfo {volume} {81}},\ \bibinfo
  {pages} {224} (\bibinfo {year} {1998})}\BibitemShut {NoStop}%
\bibitem [{\citenamefont {Geindre}\ \emph {et~al.}(1994)\citenamefont
  {Geindre}, \citenamefont {Mysyrowicz}, \citenamefont {Santos}, \citenamefont
  {Audebert}, \citenamefont {Rousse}, \citenamefont {Hamoniaux}, \citenamefont
  {Antonetti}, \citenamefont {Falli{\`{e}}s},\ and\ \citenamefont
  {Gauthier}}]{Geindre1994}%
  \BibitemOpen
  \bibfield  {author} {\bibinfo {author} {\bibfnamefont {J.~P.}\ \bibnamefont
  {Geindre}}, \bibinfo {author} {\bibfnamefont {A.}~\bibnamefont {Mysyrowicz}},
  \bibinfo {author} {\bibfnamefont {A.~Dos}\ \bibnamefont {Santos}}, \bibinfo
  {author} {\bibfnamefont {P.}~\bibnamefont {Audebert}}, \bibinfo {author}
  {\bibfnamefont {A.}~\bibnamefont {Rousse}}, \bibinfo {author} {\bibfnamefont
  {G.}~\bibnamefont {Hamoniaux}}, \bibinfo {author} {\bibfnamefont
  {A.}~\bibnamefont {Antonetti}}, \bibinfo {author} {\bibfnamefont
  {F.}~\bibnamefont {Falli{\`{e}}s}}, \ and\ \bibinfo {author} {\bibfnamefont
  {J.~C.}\ \bibnamefont {Gauthier}},\ }\bibfield  {title} {\enquote {\bibinfo
  {title} {{Frequency-domain interferometer for measuring the phase and
  amplitude of a femtosecond pulse probing a laser-produced plasma}},}\ }\href
  {\doibase 10.1364/OL.19.001997} {\bibfield  {journal} {\bibinfo  {journal}
  {Optics Letters}\ }\textbf {\bibinfo {volume} {19}},\ \bibinfo {pages} {1997}
  (\bibinfo {year} {1994})}\BibitemShut {NoStop}%
\bibitem [{\citenamefont {Bocoum}\ \emph {et~al.}(2015)\citenamefont {Bocoum},
  \citenamefont {B{\"{o}}hle}, \citenamefont {Vernier}, \citenamefont
  {Jullien}, \citenamefont {Faure},\ and\ \citenamefont
  {Lopez-Martens}}]{Bocoum2015}%
  \BibitemOpen
  \bibfield  {author} {\bibinfo {author} {\bibfnamefont {MaÃ¯mouna}\
  \bibnamefont {Bocoum}}, \bibinfo {author} {\bibfnamefont {Frederik}\
  \bibnamefont {B{\"{o}}hle}}, \bibinfo {author} {\bibfnamefont {Aline}\
  \bibnamefont {Vernier}}, \bibinfo {author} {\bibfnamefont {AurÃ©lie}\
  \bibnamefont {Jullien}}, \bibinfo {author} {\bibfnamefont {JÃ©rÃ´me}\
  \bibnamefont {Faure}}, \ and\ \bibinfo {author} {\bibfnamefont {Rodrigo}\
  \bibnamefont {Lopez-Martens}},\ }\bibfield  {title} {\enquote {\bibinfo
  {title} {{Spatial-domain interferometer for measuring plasma mirror
  expansion}},}\ }\href {\doibase 10.1364/ol.40.003009} {\bibfield  {journal}
  {\bibinfo  {journal} {Optics Letters}\ }\textbf {\bibinfo {volume} {40}},\
  \bibinfo {pages} {3009} (\bibinfo {year} {2015})}\BibitemShut {NoStop}%
\bibitem [{\citenamefont {Malvache}\ \emph {et~al.}(2013)\citenamefont
  {Malvache}, \citenamefont {Borot}, \citenamefont {Qu{\'{e}}r{\'{e}}},\ and\
  \citenamefont {Lopez-Martens}}]{Malvache2013}%
  \BibitemOpen
  \bibfield  {author} {\bibinfo {author} {\bibfnamefont {A.}~\bibnamefont
  {Malvache}}, \bibinfo {author} {\bibfnamefont {A.}~\bibnamefont {Borot}},
  \bibinfo {author} {\bibfnamefont {F.}~\bibnamefont {Qu{\'{e}}r{\'{e}}}}, \
  and\ \bibinfo {author} {\bibfnamefont {R.}~\bibnamefont {Lopez-Martens}},\
  }\bibfield  {title} {\enquote {\bibinfo {title} {{Coherent wake emission
  spectroscopy as a probe of steep plasma density profiles}},}\ }\href
  {\doibase 10.1103/PhysRevE.87.035101} {\bibfield  {journal} {\bibinfo
  {journal} {Physical Review E - Statistical, Nonlinear, and Soft Matter
  Physics}\ }\textbf {\bibinfo {volume} {87}},\ \bibinfo {pages} {035101}
  (\bibinfo {year} {2013})}\BibitemShut {NoStop}%
\bibitem [{\citenamefont {Quinn}\ \emph {et~al.}(2012)\citenamefont {Quinn},
  \citenamefont {Romagnani}, \citenamefont {Ramakrishna}, \citenamefont
  {Sarri}, \citenamefont {Dieckmann}, \citenamefont {Wilson}, \citenamefont
  {Fuchs}, \citenamefont {Lancia}, \citenamefont {Pipahl}, \citenamefont
  {Toncian}, \citenamefont {Willi}, \citenamefont {Clarke}, \citenamefont
  {Notley}, \citenamefont {MacChi},\ and\ \citenamefont
  {Borghesi}}]{Quinn2012}%
  \BibitemOpen
  \bibfield  {author} {\bibinfo {author} {\bibfnamefont {K.}~\bibnamefont
  {Quinn}}, \bibinfo {author} {\bibfnamefont {L.}~\bibnamefont {Romagnani}},
  \bibinfo {author} {\bibfnamefont {B.}~\bibnamefont {Ramakrishna}}, \bibinfo
  {author} {\bibfnamefont {G.}~\bibnamefont {Sarri}}, \bibinfo {author}
  {\bibfnamefont {M.~E.}\ \bibnamefont {Dieckmann}}, \bibinfo {author}
  {\bibfnamefont {P.~A.}\ \bibnamefont {Wilson}}, \bibinfo {author}
  {\bibfnamefont {J.}~\bibnamefont {Fuchs}}, \bibinfo {author} {\bibfnamefont
  {L.}~\bibnamefont {Lancia}}, \bibinfo {author} {\bibfnamefont
  {A.}~\bibnamefont {Pipahl}}, \bibinfo {author} {\bibfnamefont
  {T.}~\bibnamefont {Toncian}}, \bibinfo {author} {\bibfnamefont
  {O.}~\bibnamefont {Willi}}, \bibinfo {author} {\bibfnamefont {R.~J.}\
  \bibnamefont {Clarke}}, \bibinfo {author} {\bibfnamefont {M.}~\bibnamefont
  {Notley}}, \bibinfo {author} {\bibfnamefont {A.}~\bibnamefont {MacChi}}, \
  and\ \bibinfo {author} {\bibfnamefont {M.}~\bibnamefont {Borghesi}},\
  }\bibfield  {title} {\enquote {\bibinfo {title} {{Weibel-induced
  filamentation during an ultrafast laser-driven plasma expansion}},}\ }\href
  {\doibase 10.1103/PhysRevLett.108.135001} {\bibfield  {journal} {\bibinfo
  {journal} {Physical Review Letters}\ }\textbf {\bibinfo {volume} {108}},\
  \bibinfo {pages} {135001} (\bibinfo {year} {2012})}\BibitemShut {NoStop}%
\bibitem [{\citenamefont {Fletcher}\ \emph {et~al.}(2015)\citenamefont
  {Fletcher}, \citenamefont {Lee}, \citenamefont {D{\"{o}}ppner}, \citenamefont
  {Galtier}, \citenamefont {Nagler}, \citenamefont {Heimann}, \citenamefont
  {Fortmann}, \citenamefont {LePape}, \citenamefont {Ma}, \citenamefont
  {Millot}, \citenamefont {Pak}, \citenamefont {Turnbull}, \citenamefont
  {Chapman}, \citenamefont {Gericke}, \citenamefont {Vorberger}, \citenamefont
  {White}, \citenamefont {Gregori}, \citenamefont {Wei}, \citenamefont
  {Barbrel}, \citenamefont {Falcone}, \citenamefont {Kao}, \citenamefont
  {Nuhn}, \citenamefont {Welch}, \citenamefont {Zastrau}, \citenamefont
  {Neumayer}, \citenamefont {Hastings},\ and\ \citenamefont
  {Glenzer}}]{Fletcher2015}%
  \BibitemOpen
  \bibfield  {author} {\bibinfo {author} {\bibfnamefont {L.~B.}\ \bibnamefont
  {Fletcher}}, \bibinfo {author} {\bibfnamefont {H.~J.}\ \bibnamefont {Lee}},
  \bibinfo {author} {\bibfnamefont {T.}~\bibnamefont {D{\"{o}}ppner}}, \bibinfo
  {author} {\bibfnamefont {E.}~\bibnamefont {Galtier}}, \bibinfo {author}
  {\bibfnamefont {B.}~\bibnamefont {Nagler}}, \bibinfo {author} {\bibfnamefont
  {P.}~\bibnamefont {Heimann}}, \bibinfo {author} {\bibfnamefont
  {C.}~\bibnamefont {Fortmann}}, \bibinfo {author} {\bibfnamefont
  {S.}~\bibnamefont {LePape}}, \bibinfo {author} {\bibfnamefont
  {T.}~\bibnamefont {Ma}}, \bibinfo {author} {\bibfnamefont {M.}~\bibnamefont
  {Millot}}, \bibinfo {author} {\bibfnamefont {A.}~\bibnamefont {Pak}},
  \bibinfo {author} {\bibfnamefont {D.}~\bibnamefont {Turnbull}}, \bibinfo
  {author} {\bibfnamefont {D.~A.}\ \bibnamefont {Chapman}}, \bibinfo {author}
  {\bibfnamefont {D.~O.}\ \bibnamefont {Gericke}}, \bibinfo {author}
  {\bibfnamefont {J.}~\bibnamefont {Vorberger}}, \bibinfo {author}
  {\bibfnamefont {T.}~\bibnamefont {White}}, \bibinfo {author} {\bibfnamefont
  {G.}~\bibnamefont {Gregori}}, \bibinfo {author} {\bibfnamefont
  {M.}~\bibnamefont {Wei}}, \bibinfo {author} {\bibfnamefont {B.}~\bibnamefont
  {Barbrel}}, \bibinfo {author} {\bibfnamefont {R.~W.}\ \bibnamefont
  {Falcone}}, \bibinfo {author} {\bibfnamefont {C.~C.}\ \bibnamefont {Kao}},
  \bibinfo {author} {\bibfnamefont {H.}~\bibnamefont {Nuhn}}, \bibinfo {author}
  {\bibfnamefont {J.}~\bibnamefont {Welch}}, \bibinfo {author} {\bibfnamefont
  {U.}~\bibnamefont {Zastrau}}, \bibinfo {author} {\bibfnamefont
  {P.}~\bibnamefont {Neumayer}}, \bibinfo {author} {\bibfnamefont {J.~B.}\
  \bibnamefont {Hastings}}, \ and\ \bibinfo {author} {\bibfnamefont {S.~H.}\
  \bibnamefont {Glenzer}},\ }\bibfield  {title} {\enquote {\bibinfo {title}
  {{Ultrabright X-ray laser scattering for dynamic warm dense matter
  physics}},}\ }\href {\doibase 10.1038/nphoton.2015.41} {\bibfield  {journal}
  {\bibinfo  {journal} {Nature Photonics}\ }\textbf {\bibinfo {volume} {9}},\
  \bibinfo {pages} {274--279} (\bibinfo {year} {2015})}\BibitemShut {NoStop}%
\bibitem [{\citenamefont {Gorkhover}\ \emph {et~al.}(2016)\citenamefont
  {Gorkhover}, \citenamefont {Schorb}, \citenamefont {Coffee}, \citenamefont
  {Adolph}, \citenamefont {Foucar}, \citenamefont {Rupp}, \citenamefont
  {Aquila}, \citenamefont {Bozek}, \citenamefont {Epp}, \citenamefont {Erk},
  \citenamefont {Gumprecht}, \citenamefont {Holmegaard}, \citenamefont
  {Hartmann}, \citenamefont {Hartmann}, \citenamefont {Hauser}, \citenamefont
  {Holl}, \citenamefont {H{\"{o}}mke}, \citenamefont {Johnsson}, \citenamefont
  {Kimmel}, \citenamefont {K{\"{u}}hnel}, \citenamefont {Messerschmidt},
  \citenamefont {Reich}, \citenamefont {Rouz{\'{e}}e}, \citenamefont {Rudek},
  \citenamefont {Schmidt}, \citenamefont {Schulz}, \citenamefont {Soltau},
  \citenamefont {Stern}, \citenamefont {Weidenspointner}, \citenamefont
  {White}, \citenamefont {K{\"{u}}pper}, \citenamefont {Str{\"{u}}der},
  \citenamefont {Schlichting}, \citenamefont {Ullrich}, \citenamefont {Rolles},
  \citenamefont {Rudenko}, \citenamefont {M{\"{o}}ller},\ and\ \citenamefont
  {Bostedt}}]{Gorkhover2016}%
  \BibitemOpen
  \bibfield  {author} {\bibinfo {author} {\bibfnamefont {Tais}\ \bibnamefont
  {Gorkhover}}, \bibinfo {author} {\bibfnamefont {Sebastian}\ \bibnamefont
  {Schorb}}, \bibinfo {author} {\bibfnamefont {Ryan}\ \bibnamefont {Coffee}},
  \bibinfo {author} {\bibfnamefont {Marcus}\ \bibnamefont {Adolph}}, \bibinfo
  {author} {\bibfnamefont {Lutz}\ \bibnamefont {Foucar}}, \bibinfo {author}
  {\bibfnamefont {Daniela}\ \bibnamefont {Rupp}}, \bibinfo {author}
  {\bibfnamefont {Andrew}\ \bibnamefont {Aquila}}, \bibinfo {author}
  {\bibfnamefont {John~D.}\ \bibnamefont {Bozek}}, \bibinfo {author}
  {\bibfnamefont {Sascha~W.}\ \bibnamefont {Epp}}, \bibinfo {author}
  {\bibfnamefont {Benjamin}\ \bibnamefont {Erk}}, \bibinfo {author}
  {\bibfnamefont {Lars}\ \bibnamefont {Gumprecht}}, \bibinfo {author}
  {\bibfnamefont {Lotte}\ \bibnamefont {Holmegaard}}, \bibinfo {author}
  {\bibfnamefont {Andreas}\ \bibnamefont {Hartmann}}, \bibinfo {author}
  {\bibfnamefont {Robert}\ \bibnamefont {Hartmann}}, \bibinfo {author}
  {\bibfnamefont {GÃ¼nter}\ \bibnamefont {Hauser}}, \bibinfo {author}
  {\bibfnamefont {Peter}\ \bibnamefont {Holl}}, \bibinfo {author}
  {\bibfnamefont {Andre}\ \bibnamefont {H{\"{o}}mke}}, \bibinfo {author}
  {\bibfnamefont {Per}\ \bibnamefont {Johnsson}}, \bibinfo {author}
  {\bibfnamefont {Nils}\ \bibnamefont {Kimmel}}, \bibinfo {author}
  {\bibfnamefont {Kai~Uwe}\ \bibnamefont {K{\"{u}}hnel}}, \bibinfo {author}
  {\bibfnamefont {Marc}\ \bibnamefont {Messerschmidt}}, \bibinfo {author}
  {\bibfnamefont {Christian}\ \bibnamefont {Reich}}, \bibinfo {author}
  {\bibfnamefont {Arnaud}\ \bibnamefont {Rouz{\'{e}}e}}, \bibinfo {author}
  {\bibfnamefont {Benedikt}\ \bibnamefont {Rudek}}, \bibinfo {author}
  {\bibfnamefont {Carlo}\ \bibnamefont {Schmidt}}, \bibinfo {author}
  {\bibfnamefont {Joachim}\ \bibnamefont {Schulz}}, \bibinfo {author}
  {\bibfnamefont {Heike}\ \bibnamefont {Soltau}}, \bibinfo {author}
  {\bibfnamefont {Stephan}\ \bibnamefont {Stern}}, \bibinfo {author}
  {\bibfnamefont {Georg}\ \bibnamefont {Weidenspointner}}, \bibinfo {author}
  {\bibfnamefont {Bill}\ \bibnamefont {White}}, \bibinfo {author}
  {\bibfnamefont {Jochen}\ \bibnamefont {K{\"{u}}pper}}, \bibinfo {author}
  {\bibfnamefont {Lothar}\ \bibnamefont {Str{\"{u}}der}}, \bibinfo {author}
  {\bibfnamefont {Ilme}\ \bibnamefont {Schlichting}}, \bibinfo {author}
  {\bibfnamefont {Joachim}\ \bibnamefont {Ullrich}}, \bibinfo {author}
  {\bibfnamefont {Daniel}\ \bibnamefont {Rolles}}, \bibinfo {author}
  {\bibfnamefont {Artem}\ \bibnamefont {Rudenko}}, \bibinfo {author}
  {\bibfnamefont {Thomas}\ \bibnamefont {M{\"{o}}ller}}, \ and\ \bibinfo
  {author} {\bibfnamefont {Christoph}\ \bibnamefont {Bostedt}},\ }\bibfield
  {title} {\enquote {\bibinfo {title} {{Femtosecond and nanometre visualization
  of structural dynamics in superheated nanoparticles}},}\ }\href {\doibase
  10.1038/nphoton.2015.264} {\bibfield  {journal} {\bibinfo  {journal} {Nature
  Photonics}\ }\textbf {\bibinfo {volume} {10}},\ \bibinfo {pages} {93--97}
  (\bibinfo {year} {2016})}\BibitemShut {NoStop}%
\bibitem [{\citenamefont {Kluge}\ \emph {et~al.}(2017)\citenamefont {Kluge},
  \citenamefont {R{\"{o}}del}, \citenamefont {R{\"{o}}del}, \citenamefont
  {Pelka}, \citenamefont {McBride}, \citenamefont {Fletcher}, \citenamefont
  {Harmand}, \citenamefont {Krygier}, \citenamefont {Higginbotham},
  \citenamefont {Bussmann}, \citenamefont {Galtier}, \citenamefont {Gamboa},
  \citenamefont {Garcia}, \citenamefont {Garten}, \citenamefont {Glenzer},
  \citenamefont {Granados}, \citenamefont {Gutt}, \citenamefont {Lee},
  \citenamefont {Nagler}, \citenamefont {Schumaker}, \citenamefont {Tavella},
  \citenamefont {Zacharias}, \citenamefont {Schramm},\ and\ \citenamefont
  {Cowan}}]{Kluge2017}%
  \BibitemOpen
  \bibfield  {author} {\bibinfo {author} {\bibfnamefont {T.}~\bibnamefont
  {Kluge}}, \bibinfo {author} {\bibfnamefont {C.}~\bibnamefont {R{\"{o}}del}},
  \bibinfo {author} {\bibfnamefont {M.}~\bibnamefont {R{\"{o}}del}}, \bibinfo
  {author} {\bibfnamefont {A.}~\bibnamefont {Pelka}}, \bibinfo {author}
  {\bibfnamefont {E.~E.}\ \bibnamefont {McBride}}, \bibinfo {author}
  {\bibfnamefont {L.~B.}\ \bibnamefont {Fletcher}}, \bibinfo {author}
  {\bibfnamefont {M.}~\bibnamefont {Harmand}}, \bibinfo {author} {\bibfnamefont
  {A.}~\bibnamefont {Krygier}}, \bibinfo {author} {\bibfnamefont
  {A.}~\bibnamefont {Higginbotham}}, \bibinfo {author} {\bibfnamefont
  {M.}~\bibnamefont {Bussmann}}, \bibinfo {author} {\bibfnamefont
  {E.}~\bibnamefont {Galtier}}, \bibinfo {author} {\bibfnamefont
  {E.}~\bibnamefont {Gamboa}}, \bibinfo {author} {\bibfnamefont {A.~L.}\
  \bibnamefont {Garcia}}, \bibinfo {author} {\bibfnamefont {M.}~\bibnamefont
  {Garten}}, \bibinfo {author} {\bibfnamefont {S.~H.}\ \bibnamefont {Glenzer}},
  \bibinfo {author} {\bibfnamefont {E.}~\bibnamefont {Granados}}, \bibinfo
  {author} {\bibfnamefont {C.}~\bibnamefont {Gutt}}, \bibinfo {author}
  {\bibfnamefont {H.~J.}\ \bibnamefont {Lee}}, \bibinfo {author} {\bibfnamefont
  {B.}~\bibnamefont {Nagler}}, \bibinfo {author} {\bibfnamefont
  {W.}~\bibnamefont {Schumaker}}, \bibinfo {author} {\bibfnamefont
  {F.}~\bibnamefont {Tavella}}, \bibinfo {author} {\bibfnamefont
  {M.}~\bibnamefont {Zacharias}}, \bibinfo {author} {\bibfnamefont
  {U.}~\bibnamefont {Schramm}}, \ and\ \bibinfo {author} {\bibfnamefont
  {T.~E.}\ \bibnamefont {Cowan}},\ }\bibfield  {title} {\enquote {\bibinfo
  {title} {{Nanometer-scale characterization of laser-driven compression,
  shocks, and phase transitions, by x-ray scattering using free electron
  lasers}},}\ }\href {\doibase 10.1063/1.5008289} {\bibfield  {journal}
  {\bibinfo  {journal} {Physics of Plasmas}\ }\textbf {\bibinfo {volume}
  {24}},\ \bibinfo {pages} {102709} (\bibinfo {year} {2017})}\BibitemShut
  {NoStop}%
\bibitem [{\citenamefont {Kluge}\ \emph {et~al.}(2018)\citenamefont {Kluge},
  \citenamefont {R{\"{o}}del}, \citenamefont {Metzkes-Ng}, \citenamefont
  {Pelka}, \citenamefont {Garcia}, \citenamefont {Prencipe}, \citenamefont
  {Rehwald}, \citenamefont {Nakatsutsumi}, \citenamefont {McBride},
  \citenamefont {Sch{\"{o}}nherr}, \citenamefont {Garten}, \citenamefont
  {Hartley}, \citenamefont {Zacharias}, \citenamefont {Grenzer}, \citenamefont
  {Erbe}, \citenamefont {Georgiev}, \citenamefont {Galtier}, \citenamefont
  {Nam}, \citenamefont {Lee}, \citenamefont {Glenzer}, \citenamefont
  {Bussmann}, \citenamefont {Gutt}, \citenamefont {Zeil}, \citenamefont
  {R{\"{o}}del}, \citenamefont {H{\"{u}}bner}, \citenamefont {Schramm},\ and\
  \citenamefont {Cowan}}]{Kluge2018}%
  \BibitemOpen
  \bibfield  {author} {\bibinfo {author} {\bibfnamefont {Thomas}\ \bibnamefont
  {Kluge}}, \bibinfo {author} {\bibfnamefont {Melanie}\ \bibnamefont
  {R{\"{o}}del}}, \bibinfo {author} {\bibfnamefont {Josefine}\ \bibnamefont
  {Metzkes-Ng}}, \bibinfo {author} {\bibfnamefont {Alexander}\ \bibnamefont
  {Pelka}}, \bibinfo {author} {\bibfnamefont {Alejandro~Laso}\ \bibnamefont
  {Garcia}}, \bibinfo {author} {\bibfnamefont {Irene}\ \bibnamefont
  {Prencipe}}, \bibinfo {author} {\bibfnamefont {Martin}\ \bibnamefont
  {Rehwald}}, \bibinfo {author} {\bibfnamefont {Motoaki}\ \bibnamefont
  {Nakatsutsumi}}, \bibinfo {author} {\bibfnamefont {Emma~E.}\ \bibnamefont
  {McBride}}, \bibinfo {author} {\bibfnamefont {Tommy}\ \bibnamefont
  {Sch{\"{o}}nherr}}, \bibinfo {author} {\bibfnamefont {Marco}\ \bibnamefont
  {Garten}}, \bibinfo {author} {\bibfnamefont {Nicholas~J.}\ \bibnamefont
  {Hartley}}, \bibinfo {author} {\bibfnamefont {Malte}\ \bibnamefont
  {Zacharias}}, \bibinfo {author} {\bibfnamefont {JÃ¶rg}\ \bibnamefont
  {Grenzer}}, \bibinfo {author} {\bibfnamefont {Artur}\ \bibnamefont {Erbe}},
  \bibinfo {author} {\bibfnamefont {Yordan~M.}\ \bibnamefont {Georgiev}},
  \bibinfo {author} {\bibfnamefont {Eric}\ \bibnamefont {Galtier}}, \bibinfo
  {author} {\bibfnamefont {Inhyuk}\ \bibnamefont {Nam}}, \bibinfo {author}
  {\bibfnamefont {Hae~Ja}\ \bibnamefont {Lee}}, \bibinfo {author}
  {\bibfnamefont {Siegfried}\ \bibnamefont {Glenzer}}, \bibinfo {author}
  {\bibfnamefont {Michael}\ \bibnamefont {Bussmann}}, \bibinfo {author}
  {\bibfnamefont {Christian}\ \bibnamefont {Gutt}}, \bibinfo {author}
  {\bibfnamefont {Karl}\ \bibnamefont {Zeil}}, \bibinfo {author} {\bibfnamefont
  {Christian}\ \bibnamefont {R{\"{o}}del}}, \bibinfo {author} {\bibfnamefont
  {Uwe}\ \bibnamefont {H{\"{u}}bner}}, \bibinfo {author} {\bibfnamefont
  {Ulrich}\ \bibnamefont {Schramm}}, \ and\ \bibinfo {author} {\bibfnamefont
  {Thomas~E.}\ \bibnamefont {Cowan}},\ }\bibfield  {title} {\enquote {\bibinfo
  {title} {{Observation of Ultrafast Solid-Density Plasma Dynamics Using
  Femtosecond X-Ray Pulses from a Free-Electron Laser}},}\ }\href {\doibase
  10.1103/PhysRevX.8.031068} {\bibfield  {journal} {\bibinfo  {journal}
  {Physical Review X}\ }\textbf {\bibinfo {volume} {8}},\ \bibinfo {pages}
  {031068} (\bibinfo {year} {2018})}\BibitemShut {NoStop}%
\bibitem [{\citenamefont {Mo}\ \emph {et~al.}(2019)\citenamefont {Mo},
  \citenamefont {Murphy}, \citenamefont {Chen}, \citenamefont {Fossati},
  \citenamefont {Li}, \citenamefont {Wang}, \citenamefont {Wang},\ and\
  \citenamefont {Glenzer}}]{Mo2019}%
  \BibitemOpen
  \bibfield  {author} {\bibinfo {author} {\bibfnamefont {Mianzhen}\
  \bibnamefont {Mo}}, \bibinfo {author} {\bibfnamefont {Samuel}\ \bibnamefont
  {Murphy}}, \bibinfo {author} {\bibfnamefont {Zhijiang}\ \bibnamefont {Chen}},
  \bibinfo {author} {\bibfnamefont {Paul}\ \bibnamefont {Fossati}}, \bibinfo
  {author} {\bibfnamefont {Renkai}\ \bibnamefont {Li}}, \bibinfo {author}
  {\bibfnamefont {Yongqiang}\ \bibnamefont {Wang}}, \bibinfo {author}
  {\bibfnamefont {Xijie}\ \bibnamefont {Wang}}, \ and\ \bibinfo {author}
  {\bibfnamefont {Siegfried}\ \bibnamefont {Glenzer}},\ }\bibfield  {title}
  {\enquote {\bibinfo {title} {{Visualization of ultrafast melting initiated
  from radiation-driven defects in solids}},}\ }\href {\doibase
  10.1126/sciadv.aaw0392} {\bibfield  {journal} {\bibinfo  {journal} {Science
  Advances}\ }\textbf {\bibinfo {volume} {5}} (\bibinfo {year} {2019}),\
  10.1126/sciadv.aaw0392}\BibitemShut {NoStop}%
\bibitem [{\citenamefont {Gaus}\ \emph {et~al.}(2021)\citenamefont {Gaus},
  \citenamefont {Bischoff}, \citenamefont {Bussmann}, \citenamefont
  {Cunningham}, \citenamefont {Curry}, \citenamefont {E}, \citenamefont
  {Galtier}, \citenamefont {Gauthier}, \citenamefont {Laso~Garc{\'{i}}a},
  \citenamefont {Garten}, \citenamefont {Glenzer}, \citenamefont {Grenzer},
  \citenamefont {Gutt}, \citenamefont {Hartley}, \citenamefont {Huang},
  \citenamefont {H{\"{u}}bner}, \citenamefont {Kraus}, \citenamefont {Lee},
  \citenamefont {McBride}, \citenamefont {Metzkes-Ng}, \citenamefont {Nagler},
  \citenamefont {Nakatsutsumi}, \citenamefont {Nikl}, \citenamefont {Ota},
  \citenamefont {Pelka}, \citenamefont {Prencipe}, \citenamefont {Randolph},
  \citenamefont {R{\"{o}}del}, \citenamefont {Sakawa}, \citenamefont
  {Schlenvoigt}, \citenamefont {{\v{S}}m{\'{i}}d}, \citenamefont {Treffert},
  \citenamefont {Voigt}, \citenamefont {Zeil}, \citenamefont {Cowan},
  \citenamefont {Schramm}, \citenamefont {Kluge}, \citenamefont {Juncheng},
  \citenamefont {Galtier}, \citenamefont {Gauthier}, \citenamefont
  {Laso~Garc{\'{i}}a}, \citenamefont {Garten}, \citenamefont {Glenzer},
  \citenamefont {Grenzer}, \citenamefont {Gutt}, \citenamefont {Hartley},
  \citenamefont {Huang}, \citenamefont {H{\"{u}}bner}, \citenamefont {Kraus},
  \citenamefont {Lee}, \citenamefont {McBride}, \citenamefont {Metzkes-Ng},
  \citenamefont {Nagler}, \citenamefont {Nakatsutsumi}, \citenamefont {Nikl},
  \citenamefont {Ota}, \citenamefont {Pelka}, \citenamefont {Prencipe},
  \citenamefont {Randolph}, \citenamefont {R{\"{o}}del}, \citenamefont
  {Sakawa}, \citenamefont {Schlenvoigt}, \citenamefont {{\v{S}}m{\'{i}}d},
  \citenamefont {Treffert}, \citenamefont {Voigt}, \citenamefont {Zeil},
  \citenamefont {Cowan}, \citenamefont {Schramm},\ and\ \citenamefont
  {Kluge}}]{Gaus2021}%
  \BibitemOpen
  \bibfield  {author} {\bibinfo {author} {\bibfnamefont {Lennart}\ \bibnamefont
  {Gaus}}, \bibinfo {author} {\bibfnamefont {Lothar}\ \bibnamefont {Bischoff}},
  \bibinfo {author} {\bibfnamefont {Michael}\ \bibnamefont {Bussmann}},
  \bibinfo {author} {\bibfnamefont {Eric}\ \bibnamefont {Cunningham}}, \bibinfo
  {author} {\bibfnamefont {Chandra~B.}\ \bibnamefont {Curry}}, \bibinfo
  {author} {\bibfnamefont {Juncheng}\ \bibnamefont {E}}, \bibinfo {author}
  {\bibfnamefont {Eric}\ \bibnamefont {Galtier}}, \bibinfo {author}
  {\bibfnamefont {Maxence}\ \bibnamefont {Gauthier}}, \bibinfo {author}
  {\bibfnamefont {Alejandro}\ \bibnamefont {Laso~Garc{\'{i}}a}}, \bibinfo
  {author} {\bibfnamefont {Marco}\ \bibnamefont {Garten}}, \bibinfo {author}
  {\bibfnamefont {Siegfried}\ \bibnamefont {Glenzer}}, \bibinfo {author}
  {\bibfnamefont {JÃ¶rg}\ \bibnamefont {Grenzer}}, \bibinfo {author}
  {\bibfnamefont {Christian}\ \bibnamefont {Gutt}}, \bibinfo {author}
  {\bibfnamefont {Nicholas~J.}\ \bibnamefont {Hartley}}, \bibinfo {author}
  {\bibfnamefont {Lingen}\ \bibnamefont {Huang}}, \bibinfo {author}
  {\bibfnamefont {Uwe}\ \bibnamefont {H{\"{u}}bner}}, \bibinfo {author}
  {\bibfnamefont {Dominik}\ \bibnamefont {Kraus}}, \bibinfo {author}
  {\bibfnamefont {Hae~Ja}\ \bibnamefont {Lee}}, \bibinfo {author}
  {\bibfnamefont {Emma~E.}\ \bibnamefont {McBride}}, \bibinfo {author}
  {\bibfnamefont {Josefine}\ \bibnamefont {Metzkes-Ng}}, \bibinfo {author}
  {\bibfnamefont {Bob}\ \bibnamefont {Nagler}}, \bibinfo {author}
  {\bibfnamefont {Motoaki}\ \bibnamefont {Nakatsutsumi}}, \bibinfo {author}
  {\bibfnamefont {Jan}\ \bibnamefont {Nikl}}, \bibinfo {author} {\bibfnamefont
  {Masato}\ \bibnamefont {Ota}}, \bibinfo {author} {\bibfnamefont {Alexander}\
  \bibnamefont {Pelka}}, \bibinfo {author} {\bibfnamefont {Irene}\ \bibnamefont
  {Prencipe}}, \bibinfo {author} {\bibfnamefont {Lisa}\ \bibnamefont
  {Randolph}}, \bibinfo {author} {\bibfnamefont {Melanie}\ \bibnamefont
  {R{\"{o}}del}}, \bibinfo {author} {\bibfnamefont {Youichi}\ \bibnamefont
  {Sakawa}}, \bibinfo {author} {\bibfnamefont {Hans-Peter~Peter}\ \bibnamefont
  {Schlenvoigt}}, \bibinfo {author} {\bibfnamefont {Michal}\ \bibnamefont
  {{\v{S}}m{\'{i}}d}}, \bibinfo {author} {\bibfnamefont {Franziska}\
  \bibnamefont {Treffert}}, \bibinfo {author} {\bibfnamefont {Katja}\
  \bibnamefont {Voigt}}, \bibinfo {author} {\bibfnamefont {Karl}\ \bibnamefont
  {Zeil}}, \bibinfo {author} {\bibfnamefont {Thomas~E.}\ \bibnamefont {Cowan}},
  \bibinfo {author} {\bibfnamefont {Ulrich}\ \bibnamefont {Schramm}}, \bibinfo
  {author} {\bibfnamefont {Thomas}\ \bibnamefont {Kluge}}, \bibinfo {author}
  {\bibfnamefont {J.~E.}\ \bibnamefont {Juncheng}}, \bibinfo {author}
  {\bibfnamefont {Eric}\ \bibnamefont {Galtier}}, \bibinfo {author}
  {\bibfnamefont {Maxence}\ \bibnamefont {Gauthier}}, \bibinfo {author}
  {\bibfnamefont {Alejandro}\ \bibnamefont {Laso~Garc{\'{i}}a}}, \bibinfo
  {author} {\bibfnamefont {Marco}\ \bibnamefont {Garten}}, \bibinfo {author}
  {\bibfnamefont {Siegfried}\ \bibnamefont {Glenzer}}, \bibinfo {author}
  {\bibfnamefont {JÃ¶rg}\ \bibnamefont {Grenzer}}, \bibinfo {author}
  {\bibfnamefont {Christian}\ \bibnamefont {Gutt}}, \bibinfo {author}
  {\bibfnamefont {Nicholas~J.}\ \bibnamefont {Hartley}}, \bibinfo {author}
  {\bibfnamefont {Lingen}\ \bibnamefont {Huang}}, \bibinfo {author}
  {\bibfnamefont {Uwe}\ \bibnamefont {H{\"{u}}bner}}, \bibinfo {author}
  {\bibfnamefont {Dominik}\ \bibnamefont {Kraus}}, \bibinfo {author}
  {\bibfnamefont {Hae~Ja}\ \bibnamefont {Lee}}, \bibinfo {author}
  {\bibfnamefont {Emma~E.}\ \bibnamefont {McBride}}, \bibinfo {author}
  {\bibfnamefont {Josefine}\ \bibnamefont {Metzkes-Ng}}, \bibinfo {author}
  {\bibfnamefont {Bob}\ \bibnamefont {Nagler}}, \bibinfo {author}
  {\bibfnamefont {Motoaki}\ \bibnamefont {Nakatsutsumi}}, \bibinfo {author}
  {\bibfnamefont {Jan}\ \bibnamefont {Nikl}}, \bibinfo {author} {\bibfnamefont
  {Masato}\ \bibnamefont {Ota}}, \bibinfo {author} {\bibfnamefont {Alexander}\
  \bibnamefont {Pelka}}, \bibinfo {author} {\bibfnamefont {Irene}\ \bibnamefont
  {Prencipe}}, \bibinfo {author} {\bibfnamefont {Lisa}\ \bibnamefont
  {Randolph}}, \bibinfo {author} {\bibfnamefont {Melanie}\ \bibnamefont
  {R{\"{o}}del}}, \bibinfo {author} {\bibfnamefont {Youichi}\ \bibnamefont
  {Sakawa}}, \bibinfo {author} {\bibfnamefont {Hans-Peter~Peter}\ \bibnamefont
  {Schlenvoigt}}, \bibinfo {author} {\bibfnamefont {Michal}\ \bibnamefont
  {{\v{S}}m{\'{i}}d}}, \bibinfo {author} {\bibfnamefont {Franziska}\
  \bibnamefont {Treffert}}, \bibinfo {author} {\bibfnamefont {Katja}\
  \bibnamefont {Voigt}}, \bibinfo {author} {\bibfnamefont {Karl}\ \bibnamefont
  {Zeil}}, \bibinfo {author} {\bibfnamefont {Thomas~E.}\ \bibnamefont {Cowan}},
  \bibinfo {author} {\bibfnamefont {Ulrich}\ \bibnamefont {Schramm}}, \ and\
  \bibinfo {author} {\bibfnamefont {Thomas}\ \bibnamefont {Kluge}},\ }\bibfield
   {title} {\enquote {\bibinfo {title} {{Probing ultrafast laser plasma
  processes inside solids with resonant small-angle x-ray scattering}},}\
  }\href {\doibase 10.1103/PhysRevResearch.3.043194} {\bibfield  {journal}
  {\bibinfo  {journal} {Physical Review Research}\ }\textbf {\bibinfo {volume}
  {3}},\ \bibinfo {pages} {043194} (\bibinfo {year} {2021})}\BibitemShut
  {NoStop}%
\bibitem [{\citenamefont {Davidson}(1972)}]{Davidson1972}%
  \BibitemOpen
  \bibfield  {author} {\bibinfo {author} {\bibfnamefont {Ronald~C.}\
  \bibnamefont {Davidson}},\ }\bibfield  {title} {\enquote {\bibinfo {title}
  {{Nonlinear Development of Electromagnetic Instabilities in Anisotropic
  Plasmas}},}\ }\href {\doibase 10.1063/1.1693910} {\bibfield  {journal}
  {\bibinfo  {journal} {Physics of Fluids}\ }\textbf {\bibinfo {volume} {15}},\
  \bibinfo {pages} {317} (\bibinfo {year} {1972})}\BibitemShut {NoStop}%
\bibitem [{\citenamefont {Tatarakis}\ \emph {et~al.}(2003)\citenamefont
  {Tatarakis}, \citenamefont {Beg}, \citenamefont {Clark}, \citenamefont
  {Dangor}, \citenamefont {Edwards}, \citenamefont {Evans}, \citenamefont
  {Goldsack}, \citenamefont {Ledingham}, \citenamefont {Norreys}, \citenamefont
  {Sinclair}, \citenamefont {Wei}, \citenamefont {Zepf},\ and\ \citenamefont
  {Krushelnick}}]{Tatarakis2003}%
  \BibitemOpen
  \bibfield  {author} {\bibinfo {author} {\bibfnamefont {M.}~\bibnamefont
  {Tatarakis}}, \bibinfo {author} {\bibfnamefont {F.~N.}\ \bibnamefont {Beg}},
  \bibinfo {author} {\bibfnamefont {E.~L.}\ \bibnamefont {Clark}}, \bibinfo
  {author} {\bibfnamefont {A.~E.}\ \bibnamefont {Dangor}}, \bibinfo {author}
  {\bibfnamefont {R.~D.}\ \bibnamefont {Edwards}}, \bibinfo {author}
  {\bibfnamefont {R.~G.}\ \bibnamefont {Evans}}, \bibinfo {author}
  {\bibfnamefont {T.~J.}\ \bibnamefont {Goldsack}}, \bibinfo {author}
  {\bibfnamefont {K.~W.D.}\ \bibnamefont {Ledingham}}, \bibinfo {author}
  {\bibfnamefont {P.~A.}\ \bibnamefont {Norreys}}, \bibinfo {author}
  {\bibfnamefont {M.~A.}\ \bibnamefont {Sinclair}}, \bibinfo {author}
  {\bibfnamefont {M.~S.}\ \bibnamefont {Wei}}, \bibinfo {author} {\bibfnamefont
  {M.}~\bibnamefont {Zepf}}, \ and\ \bibinfo {author} {\bibfnamefont
  {K.}~\bibnamefont {Krushelnick}},\ }\bibfield  {title} {\enquote {\bibinfo
  {title} {{Propagation Instabilities of High-Intensity Laser-Produced Electron
  Beams}},}\ }\href {\doibase 10.1103/PhysRevLett.90.175001} {\bibfield
  {journal} {\bibinfo  {journal} {Physical Review Letters}\ }\textbf {\bibinfo
  {volume} {90}},\ \bibinfo {pages} {4} (\bibinfo {year} {2003})}\BibitemShut
  {NoStop}%
\bibitem [{\citenamefont {Bret}\ \emph
  {et~al.}(2005{\natexlab{a}})\citenamefont {Bret}, \citenamefont {Firpo},\
  and\ \citenamefont {Deutsch}}]{Bret2005}%
  \BibitemOpen
  \bibfield  {author} {\bibinfo {author} {\bibfnamefont {A.}~\bibnamefont
  {Bret}}, \bibinfo {author} {\bibfnamefont {M.-C.}\ \bibnamefont {Firpo}}, \
  and\ \bibinfo {author} {\bibfnamefont {C.}~\bibnamefont {Deutsch}},\
  }\bibfield  {title} {\enquote {\bibinfo {title} {{Characterization of the
  Initial Filamentation of a Relativistic Electron Beam Passing through a
  Plasma}},}\ }\href {\doibase 10.1103/PhysRevLett.94.115002} {\bibfield
  {journal} {\bibinfo  {journal} {Physical Review Letters}\ }\textbf {\bibinfo
  {volume} {94}},\ \bibinfo {pages} {115002} (\bibinfo {year}
  {2005}{\natexlab{a}})}\BibitemShut {NoStop}%
\bibitem [{\citenamefont {Bret}\ \emph
  {et~al.}(2005{\natexlab{b}})\citenamefont {Bret}, \citenamefont {Firpo},\
  and\ \citenamefont {Deutsch}}]{Bret2005a}%
  \BibitemOpen
  \bibfield  {author} {\bibinfo {author} {\bibfnamefont {A.}~\bibnamefont
  {Bret}}, \bibinfo {author} {\bibfnamefont {M.-C.}\ \bibnamefont {Firpo}}, \
  and\ \bibinfo {author} {\bibfnamefont {C.}~\bibnamefont {Deutsch}},\
  }\bibfield  {title} {\enquote {\bibinfo {title} {{Electromagnetic
  instabilities for relativistic beam-plasma interaction in whole <math
  display="inline"> <mi>k</mi> </math> space: Nonrelativistic beam and plasma
  temperature effects}},}\ }\href {\doibase 10.1103/PhysRevE.72.016403}
  {\bibfield  {journal} {\bibinfo  {journal} {Physical Review E}\ }\textbf
  {\bibinfo {volume} {72}},\ \bibinfo {pages} {016403} (\bibinfo {year}
  {2005}{\natexlab{b}})}\BibitemShut {NoStop}%
\bibitem [{\citenamefont {Cottrill}\ \emph {et~al.}(2008)\citenamefont
  {Cottrill}, \citenamefont {Langdon}, \citenamefont {Lasinski}, \citenamefont
  {Lund}, \citenamefont {Molvig}, \citenamefont {Tabak}, \citenamefont {Town},\
  and\ \citenamefont {Williams}}]{Cottrill2008}%
  \BibitemOpen
  \bibfield  {author} {\bibinfo {author} {\bibfnamefont {L.~A.}\ \bibnamefont
  {Cottrill}}, \bibinfo {author} {\bibfnamefont {A.~B.}\ \bibnamefont
  {Langdon}}, \bibinfo {author} {\bibfnamefont {B.~F.}\ \bibnamefont
  {Lasinski}}, \bibinfo {author} {\bibfnamefont {S.~M.}\ \bibnamefont {Lund}},
  \bibinfo {author} {\bibfnamefont {K.}~\bibnamefont {Molvig}}, \bibinfo
  {author} {\bibfnamefont {M.}~\bibnamefont {Tabak}}, \bibinfo {author}
  {\bibfnamefont {R.~P.~J.}\ \bibnamefont {Town}}, \ and\ \bibinfo {author}
  {\bibfnamefont {E.~A.}\ \bibnamefont {Williams}},\ }\bibfield  {title}
  {\enquote {\bibinfo {title} {{Kinetic and collisional effects on the linear
  evolution of fast ignition relevant beam instabilities}},}\ }\href {\doibase
  10.1063/1.2953816} {\bibfield  {journal} {\bibinfo  {journal} {Physics of
  Plasmas}\ }\textbf {\bibinfo {volume} {15}} (\bibinfo {year} {2008}),\
  10.1063/1.2953816}\BibitemShut {NoStop}%
\bibitem [{\citenamefont {Bret}\ \emph {et~al.}(2010)\citenamefont {Bret},
  \citenamefont {Gremillet},\ and\ \citenamefont {B{\'{e}}nisti}}]{Bret2010}%
  \BibitemOpen
  \bibfield  {author} {\bibinfo {author} {\bibfnamefont {A.}~\bibnamefont
  {Bret}}, \bibinfo {author} {\bibfnamefont {L.}~\bibnamefont {Gremillet}}, \
  and\ \bibinfo {author} {\bibfnamefont {D.}~\bibnamefont {B{\'{e}}nisti}},\
  }\bibfield  {title} {\enquote {\bibinfo {title} {{Exact relativistic kinetic
  theory of the full unstable spectrum of an electron-beamâ€“plasma system with
  Maxwell-J{\"{u}}ttner distribution functions}},}\ }\href {\doibase
  10.1103/PhysRevE.81.036402} {\bibfield  {journal} {\bibinfo  {journal}
  {Physical Review E}\ }\textbf {\bibinfo {volume} {81}},\ \bibinfo {pages}
  {036402} (\bibinfo {year} {2010})}\BibitemShut {NoStop}%
\bibitem [{\citenamefont {Bret}\ \emph {et~al.}(2014)\citenamefont {Bret},
  \citenamefont {Stockem}, \citenamefont {Narayan},\ and\ \citenamefont
  {Silva}}]{bret2014}%
  \BibitemOpen
  \bibfield  {author} {\bibinfo {author} {\bibfnamefont {A.}~\bibnamefont
  {Bret}}, \bibinfo {author} {\bibfnamefont {A.}~\bibnamefont {Stockem}},
  \bibinfo {author} {\bibfnamefont {R.}~\bibnamefont {Narayan}}, \ and\
  \bibinfo {author} {\bibfnamefont {L.~O.}\ \bibnamefont {Silva}},\ }\bibfield
  {title} {\enquote {\bibinfo {title} {{Collisionless Weibel shocks: Full
  formation mechanism and timing}},}\ }\href {\doibase 10.1063/1.4886121}
  {\bibfield  {journal} {\bibinfo  {journal} {Physics of Plasmas}\ }\textbf
  {\bibinfo {volume} {21}},\ \bibinfo {pages} {072301} (\bibinfo {year}
  {2014})}\BibitemShut {NoStop}%
\bibitem [{\citenamefont {Sherlock}\ \emph {et~al.}(2014)\citenamefont
  {Sherlock}, \citenamefont {Hill}, \citenamefont {Evans}, \citenamefont
  {Rose},\ and\ \citenamefont {Rozmus}}]{Sherlock2014}%
  \BibitemOpen
  \bibfield  {author} {\bibinfo {author} {\bibfnamefont {M.}~\bibnamefont
  {Sherlock}}, \bibinfo {author} {\bibfnamefont {E.~G.}\ \bibnamefont {Hill}},
  \bibinfo {author} {\bibfnamefont {R.~G.}\ \bibnamefont {Evans}}, \bibinfo
  {author} {\bibfnamefont {S.~J.}\ \bibnamefont {Rose}}, \ and\ \bibinfo
  {author} {\bibfnamefont {W.}~\bibnamefont {Rozmus}},\ }\bibfield  {title}
  {\enquote {\bibinfo {title} {{In-depth plasma-wave heating of dense plasma
  irradiated by short laser pulses}},}\ }\href {\doibase
  10.1103/PhysRevLett.113.255001} {\bibfield  {journal} {\bibinfo  {journal}
  {Physical Review Letters}\ }\textbf {\bibinfo {volume} {113}},\ \bibinfo
  {pages} {255001} (\bibinfo {year} {2014})}\BibitemShut {NoStop}%
\bibitem [{\citenamefont {Macchi}(2018)}]{Macchi2018}%
  \BibitemOpen
  \bibfield  {author} {\bibinfo {author} {\bibfnamefont {A.}~\bibnamefont
  {Macchi}},\ }\bibfield  {title} {\enquote {\bibinfo {title} {{Surface
  plasmons in superintense laser-solid interactions}},}\ }\href {\doibase
  10.1063/1.5013321} {\bibfield  {journal} {\bibinfo  {journal} {Physics of
  Plasmas}\ }\textbf {\bibinfo {volume} {25}} (\bibinfo {year} {2018}),\
  10.1063/1.5013321}\BibitemShut {NoStop}%
\bibitem [{\citenamefont {Fedeli}\ \emph {et~al.}(2016)\citenamefont {Fedeli},
  \citenamefont {Sgattoni}, \citenamefont {Cantono}, \citenamefont {Garzella},
  \citenamefont {R{\'{e}}au}, \citenamefont {Prencipe}, \citenamefont
  {Passoni}, \citenamefont {Raynaud}, \citenamefont {Kv{\v{e}}toÅˆ},
  \citenamefont {Proska}, \citenamefont {Macchi},\ and\ \citenamefont
  {Ceccotti}}]{Fedeli2016}%
  \BibitemOpen
  \bibfield  {author} {\bibinfo {author} {\bibfnamefont {L.}~\bibnamefont
  {Fedeli}}, \bibinfo {author} {\bibfnamefont {A.}~\bibnamefont {Sgattoni}},
  \bibinfo {author} {\bibfnamefont {G.}~\bibnamefont {Cantono}}, \bibinfo
  {author} {\bibfnamefont {D.}~\bibnamefont {Garzella}}, \bibinfo {author}
  {\bibfnamefont {F.}~\bibnamefont {R{\'{e}}au}}, \bibinfo {author}
  {\bibfnamefont {I.}~\bibnamefont {Prencipe}}, \bibinfo {author}
  {\bibfnamefont {M.}~\bibnamefont {Passoni}}, \bibinfo {author} {\bibfnamefont
  {M.}~\bibnamefont {Raynaud}}, \bibinfo {author} {\bibfnamefont
  {M.}~\bibnamefont {Kv{\v{e}}toÅˆ}}, \bibinfo {author} {\bibfnamefont
  {J.}~\bibnamefont {Proska}}, \bibinfo {author} {\bibfnamefont
  {A.}~\bibnamefont {Macchi}}, \ and\ \bibinfo {author} {\bibfnamefont
  {T.}~\bibnamefont {Ceccotti}},\ }\bibfield  {title} {\enquote {\bibinfo
  {title} {{Electron Acceleration by Relativistic Surface Plasmons in
  Laser-Grating Interaction}},}\ }\href {\doibase
  10.1103/PhysRevLett.116.015001} {\bibfield  {journal} {\bibinfo  {journal}
  {Physical Review Letters}\ }\textbf {\bibinfo {volume} {116}},\ \bibinfo
  {pages} {015001} (\bibinfo {year} {2016})}\BibitemShut {NoStop}%
\bibitem [{\citenamefont {Laso~Garcia}\ \emph {et~al.}(2021)\citenamefont
  {Laso~Garcia}, \citenamefont {H{\"{o}}ppner}, \citenamefont {Pelka},
  \citenamefont {B{\"{a}}htz}, \citenamefont {Brambrink}, \citenamefont
  {Di~Dio~Cafiso}, \citenamefont {Dreyer}, \citenamefont {G{\"{o}}de},
  \citenamefont {Hassan}, \citenamefont {Kluge}, \citenamefont {Liu},
  \citenamefont {Makita}, \citenamefont {M{\"{o}}ller}, \citenamefont
  {Nakatsutsumi}, \citenamefont {Preston}, \citenamefont {Priebe},
  \citenamefont {Schlenvoigt}, \citenamefont {Schwinkendorf}, \citenamefont
  {{\v{S}}m{\'{i}}d}, \citenamefont {Talposi}, \citenamefont {Toncian},
  \citenamefont {Zastrau}, \citenamefont {Schramm}, \citenamefont {Cowan},\
  and\ \citenamefont {Toncian}}]{lasogarcia2021}%
  \BibitemOpen
  \bibfield  {author} {\bibinfo {author} {\bibfnamefont {A.}~\bibnamefont
  {Laso~Garcia}}, \bibinfo {author} {\bibfnamefont {H.}~\bibnamefont
  {H{\"{o}}ppner}}, \bibinfo {author} {\bibfnamefont {A.}~\bibnamefont
  {Pelka}}, \bibinfo {author} {\bibfnamefont {C.}~\bibnamefont {B{\"{a}}htz}},
  \bibinfo {author} {\bibfnamefont {E.}~\bibnamefont {Brambrink}}, \bibinfo
  {author} {\bibfnamefont {S.}~\bibnamefont {Di~Dio~Cafiso}}, \bibinfo {author}
  {\bibfnamefont {J.}~\bibnamefont {Dreyer}}, \bibinfo {author} {\bibfnamefont
  {S.}~\bibnamefont {G{\"{o}}de}}, \bibinfo {author} {\bibfnamefont
  {M.}~\bibnamefont {Hassan}}, \bibinfo {author} {\bibfnamefont
  {T.}~\bibnamefont {Kluge}}, \bibinfo {author} {\bibfnamefont
  {J.}~\bibnamefont {Liu}}, \bibinfo {author} {\bibfnamefont {M.}~\bibnamefont
  {Makita}}, \bibinfo {author} {\bibfnamefont {D.}~\bibnamefont
  {M{\"{o}}ller}}, \bibinfo {author} {\bibfnamefont {M.}~\bibnamefont
  {Nakatsutsumi}}, \bibinfo {author} {\bibfnamefont {T.~R.}\ \bibnamefont
  {Preston}}, \bibinfo {author} {\bibfnamefont {G.}~\bibnamefont {Priebe}},
  \bibinfo {author} {\bibfnamefont {H.-P.~P.}\ \bibnamefont {Schlenvoigt}},
  \bibinfo {author} {\bibfnamefont {J.-P.~P.}\ \bibnamefont {Schwinkendorf}},
  \bibinfo {author} {\bibfnamefont {M.}~\bibnamefont {{\v{S}}m{\'{i}}d}},
  \bibinfo {author} {\bibfnamefont {A.-M.~M.}\ \bibnamefont {Talposi}},
  \bibinfo {author} {\bibfnamefont {M.}~\bibnamefont {Toncian}}, \bibinfo
  {author} {\bibfnamefont {U.}~\bibnamefont {Zastrau}}, \bibinfo {author}
  {\bibfnamefont {U.}~\bibnamefont {Schramm}}, \bibinfo {author} {\bibfnamefont
  {T.~E.}\ \bibnamefont {Cowan}}, \ and\ \bibinfo {author} {\bibfnamefont
  {T.}~\bibnamefont {Toncian}},\ }\bibfield  {title} {\enquote {\bibinfo
  {title} {{ReLaX: the Helmholtz International Beamline for Extreme Fields
  high-intensity short-pulse laser driver for relativistic laserâ€“matter
  interaction and strong-field science using the high energy density instrument
  at the European X-ray free electron laser fa}},}\ }\href {\doibase
  10.1017/hpl.2021.47} {\bibfield  {journal} {\bibinfo  {journal} {High Power
  Laser Science and Engineering}\ }\textbf {\bibinfo {volume} {9}},\ \bibinfo
  {pages} {E59} (\bibinfo {year} {2021})}\BibitemShut {NoStop}%
\bibitem [{\citenamefont {{HIBEF Consortium}}()}]{HIBEFConsortium}%
  \BibitemOpen
  \bibfield  {author} {\bibinfo {author} {\bibnamefont {{HIBEF Consortium}}},\
  }\href@noop {} {\enquote {\bibinfo {title} {{http://www.hibef.eu}},}\
  }\BibitemShut {NoStop}%
\bibitem [{\citenamefont {Nakatsutsumi}\ and\ \citenamefont
  {Tschentscher}(2013)}]{Tschentscher2013}%
  \BibitemOpen
  \bibfield  {author} {\bibinfo {author} {\bibfnamefont {Motoaki}\ \bibnamefont
  {Nakatsutsumi}}\ and\ \bibinfo {author} {\bibfnamefont {Thomas}\ \bibnamefont
  {Tschentscher}},\ }\href {\doibase 10.3204/XFEL.EU/TR-2014-001} {\emph
  {\bibinfo {title} {Cdr}}},\ \bibinfo {type} {Tech. Rep.}\ (\bibinfo
  {institution} {European X-Ray Free-Electron Laser Facility GmbH},\ \bibinfo
  {year} {2013})\BibitemShut {NoStop}%
\bibitem [{\citenamefont {Zastrau}\ \emph {et~al.}(2021)\citenamefont
  {Zastrau}, \citenamefont {Appel}, \citenamefont {Baehtz}, \citenamefont
  {Baehr}, \citenamefont {Batchelor}, \citenamefont {Bergh{\"{a}}user},
  \citenamefont {Banjafar}, \citenamefont {Brambrink}, \citenamefont
  {Cerantola}, \citenamefont {Cowan}, \citenamefont {Damker}, \citenamefont
  {Dietrich}, \citenamefont {Di~Dio~Cafiso}, \citenamefont {Dreyer},
  \citenamefont {Engel}, \citenamefont {Feldmann}, \citenamefont {Findeisen},
  \citenamefont {Foese}, \citenamefont {Fulla-Marsa}, \citenamefont
  {G{\"{o}}de}, \citenamefont {Hassan}, \citenamefont {Hauser}, \citenamefont
  {Herrmannsd{\"{o}}rfer}, \citenamefont {H{\"{o}}ppner}, \citenamefont {Kaa},
  \citenamefont {Kaever}, \citenamefont {Kn{\"{o}}fel}, \citenamefont
  {Konopkov{\'{a}}}, \citenamefont {Garc{\'{i}}a}, \citenamefont {Liermann},
  \citenamefont {Mainberger}, \citenamefont {Makit}, \citenamefont {Martens},
  \citenamefont {McBride}, \citenamefont {M{\"{o}}ller}, \citenamefont
  {Nakatsutsumi}, \citenamefont {Pelka}, \citenamefont {Plueckthun},
  \citenamefont {Prescher}, \citenamefont {Preston}, \citenamefont
  {R{\"{o}}per}, \citenamefont {Schmidt}, \citenamefont {Seidel}, \citenamefont
  {Schwinkendorf}, \citenamefont {Schoelmerich}, \citenamefont {Schramm},
  \citenamefont {Schropp}, \citenamefont {Strohm}, \citenamefont {Sukharnikov},
  \citenamefont {Talkovski}, \citenamefont {Thorpe}, \citenamefont {Toncian},
  \citenamefont {Toncian}, \citenamefont {Wollenweber}, \citenamefont
  {Yamamoto},\ and\ \citenamefont {Tschentscher}}]{Zastrau2021}%
  \BibitemOpen
  \bibfield  {author} {\bibinfo {author} {\bibfnamefont {Ulf}\ \bibnamefont
  {Zastrau}}, \bibinfo {author} {\bibfnamefont {Karen}\ \bibnamefont {Appel}},
  \bibinfo {author} {\bibfnamefont {Carsten}\ \bibnamefont {Baehtz}}, \bibinfo
  {author} {\bibfnamefont {Oliver}\ \bibnamefont {Baehr}}, \bibinfo {author}
  {\bibfnamefont {Lewis}\ \bibnamefont {Batchelor}}, \bibinfo {author}
  {\bibfnamefont {Andreas}\ \bibnamefont {Bergh{\"{a}}user}}, \bibinfo {author}
  {\bibfnamefont {Mohammadreza}\ \bibnamefont {Banjafar}}, \bibinfo {author}
  {\bibfnamefont {Erik}\ \bibnamefont {Brambrink}}, \bibinfo {author}
  {\bibfnamefont {Valerio}\ \bibnamefont {Cerantola}}, \bibinfo {author}
  {\bibfnamefont {Thomas~E.}\ \bibnamefont {Cowan}}, \bibinfo {author}
  {\bibfnamefont {Horst}\ \bibnamefont {Damker}}, \bibinfo {author}
  {\bibfnamefont {Steffen}\ \bibnamefont {Dietrich}}, \bibinfo {author}
  {\bibfnamefont {Samuele}\ \bibnamefont {Di~Dio~Cafiso}}, \bibinfo {author}
  {\bibfnamefont {JÃ¶rn}\ \bibnamefont {Dreyer}}, \bibinfo {author}
  {\bibfnamefont {Hans~Olaf}\ \bibnamefont {Engel}}, \bibinfo {author}
  {\bibfnamefont {Thomas}\ \bibnamefont {Feldmann}}, \bibinfo {author}
  {\bibfnamefont {Stefan}\ \bibnamefont {Findeisen}}, \bibinfo {author}
  {\bibfnamefont {Manon}\ \bibnamefont {Foese}}, \bibinfo {author}
  {\bibfnamefont {Daniel}\ \bibnamefont {Fulla-Marsa}}, \bibinfo {author}
  {\bibfnamefont {Sebastian}\ \bibnamefont {G{\"{o}}de}}, \bibinfo {author}
  {\bibfnamefont {Mohammed}\ \bibnamefont {Hassan}}, \bibinfo {author}
  {\bibfnamefont {Jens}\ \bibnamefont {Hauser}}, \bibinfo {author}
  {\bibfnamefont {Thomas}\ \bibnamefont {Herrmannsd{\"{o}}rfer}}, \bibinfo
  {author} {\bibfnamefont {Hauke}\ \bibnamefont {H{\"{o}}ppner}}, \bibinfo
  {author} {\bibfnamefont {Johannes}\ \bibnamefont {Kaa}}, \bibinfo {author}
  {\bibfnamefont {Peter}\ \bibnamefont {Kaever}}, \bibinfo {author}
  {\bibfnamefont {Klaus}\ \bibnamefont {Kn{\"{o}}fel}}, \bibinfo {author}
  {\bibfnamefont {Zuzana}\ \bibnamefont {Konopkov{\'{a}}}}, \bibinfo {author}
  {\bibfnamefont {Alejandro~Laso}\ \bibnamefont {Garc{\'{i}}a}}, \bibinfo
  {author} {\bibfnamefont {Hanns~Peter}\ \bibnamefont {Liermann}}, \bibinfo
  {author} {\bibfnamefont {Jona}\ \bibnamefont {Mainberger}}, \bibinfo {author}
  {\bibfnamefont {Mikako}\ \bibnamefont {Makit}}, \bibinfo {author}
  {\bibfnamefont {Eike~Christian}\ \bibnamefont {Martens}}, \bibinfo {author}
  {\bibfnamefont {Emma~E.}\ \bibnamefont {McBride}}, \bibinfo {author}
  {\bibfnamefont {Dominik}\ \bibnamefont {M{\"{o}}ller}}, \bibinfo {author}
  {\bibfnamefont {Motoaki}\ \bibnamefont {Nakatsutsumi}}, \bibinfo {author}
  {\bibfnamefont {Alexander}\ \bibnamefont {Pelka}}, \bibinfo {author}
  {\bibfnamefont {Christian}\ \bibnamefont {Plueckthun}}, \bibinfo {author}
  {\bibfnamefont {Clemens}\ \bibnamefont {Prescher}}, \bibinfo {author}
  {\bibfnamefont {Thomas~R.}\ \bibnamefont {Preston}}, \bibinfo {author}
  {\bibfnamefont {Michael}\ \bibnamefont {R{\"{o}}per}}, \bibinfo {author}
  {\bibfnamefont {Andreas}\ \bibnamefont {Schmidt}}, \bibinfo {author}
  {\bibfnamefont {Wolfgang}\ \bibnamefont {Seidel}}, \bibinfo {author}
  {\bibfnamefont {Jan~Patrick}\ \bibnamefont {Schwinkendorf}}, \bibinfo
  {author} {\bibfnamefont {Markus~O.}\ \bibnamefont {Schoelmerich}}, \bibinfo
  {author} {\bibfnamefont {Ulrich}\ \bibnamefont {Schramm}}, \bibinfo {author}
  {\bibfnamefont {Andreas}\ \bibnamefont {Schropp}}, \bibinfo {author}
  {\bibfnamefont {Cornelius}\ \bibnamefont {Strohm}}, \bibinfo {author}
  {\bibfnamefont {Konstantin}\ \bibnamefont {Sukharnikov}}, \bibinfo {author}
  {\bibfnamefont {Peter}\ \bibnamefont {Talkovski}}, \bibinfo {author}
  {\bibfnamefont {Ian}\ \bibnamefont {Thorpe}}, \bibinfo {author}
  {\bibfnamefont {Monika}\ \bibnamefont {Toncian}}, \bibinfo {author}
  {\bibfnamefont {Toma}\ \bibnamefont {Toncian}}, \bibinfo {author}
  {\bibfnamefont {Lennart}\ \bibnamefont {Wollenweber}}, \bibinfo {author}
  {\bibfnamefont {Shingo}\ \bibnamefont {Yamamoto}}, \ and\ \bibinfo {author}
  {\bibfnamefont {Thomas}\ \bibnamefont {Tschentscher}},\ }\bibfield  {title}
  {\enquote {\bibinfo {title} {{The High Energy Density Scientific Instrument
  at the European XFEL}},}\ }\href {\doibase 10.1107/S1600577521007335}
  {\bibfield  {journal} {\bibinfo  {journal} {urn:issn:1600-5775}\ }\textbf
  {\bibinfo {volume} {28}},\ \bibinfo {pages} {1393--1416} (\bibinfo {year}
  {2021})}\BibitemShut {NoStop}%
\bibitem [{\citenamefont {{\v{S}}m{\'{i}}d}\ \emph {et~al.}(2020)\citenamefont
  {{\v{S}}m{\'{i}}d}, \citenamefont {Baehtz}, \citenamefont {Pelka},
  \citenamefont {Laso~Garc{\'{i}}a}, \citenamefont {G{\"{o}}de}, \citenamefont
  {Grenzer}, \citenamefont {Kluge}, \citenamefont {Konopkova}, \citenamefont
  {Makita}, \citenamefont {Prencipe}, \citenamefont {Preston}, \citenamefont
  {R{\"{o}}del},\ and\ \citenamefont {Cowan}}]{Smid2020}%
  \BibitemOpen
  \bibfield  {author} {\bibinfo {author} {\bibfnamefont {M.}~\bibnamefont
  {{\v{S}}m{\'{i}}d}}, \bibinfo {author} {\bibfnamefont {C.}~\bibnamefont
  {Baehtz}}, \bibinfo {author} {\bibfnamefont {A.}~\bibnamefont {Pelka}},
  \bibinfo {author} {\bibfnamefont {A.}~\bibnamefont {Laso~Garc{\'{i}}a}},
  \bibinfo {author} {\bibfnamefont {S.}~\bibnamefont {G{\"{o}}de}}, \bibinfo
  {author} {\bibfnamefont {J.}~\bibnamefont {Grenzer}}, \bibinfo {author}
  {\bibfnamefont {T.}~\bibnamefont {Kluge}}, \bibinfo {author} {\bibfnamefont
  {Z.}~\bibnamefont {Konopkova}}, \bibinfo {author} {\bibfnamefont
  {M.}~\bibnamefont {Makita}}, \bibinfo {author} {\bibfnamefont
  {I.}~\bibnamefont {Prencipe}}, \bibinfo {author} {\bibfnamefont {T.~R.}\
  \bibnamefont {Preston}}, \bibinfo {author} {\bibfnamefont {M.}~\bibnamefont
  {R{\"{o}}del}}, \ and\ \bibinfo {author} {\bibfnamefont {T.~E.}\ \bibnamefont
  {Cowan}},\ }\bibfield  {title} {\enquote {\bibinfo {title} {{Mirror to
  measure small angle x-ray scattering signal in high energy density
  experiments}},}\ }\href {\doibase 10.1063/5.0021691} {\bibfield  {journal}
  {\bibinfo  {journal} {Review of Scientific Instruments}\ }\textbf {\bibinfo
  {volume} {91}},\ \bibinfo {pages} {123501} (\bibinfo {year}
  {2020})}\BibitemShut {NoStop}%
\bibitem [{\citenamefont {Kluge}\ \emph {et~al.}(2014)\citenamefont {Kluge},
  \citenamefont {Gutt}, \citenamefont {Huang}, \citenamefont {Metzkes},
  \citenamefont {Schramm}, \citenamefont {Bussmann},\ and\ \citenamefont
  {Cowan}}]{Kluge2014}%
  \BibitemOpen
  \bibfield  {author} {\bibinfo {author} {\bibfnamefont {T.}~\bibnamefont
  {Kluge}}, \bibinfo {author} {\bibfnamefont {C.}~\bibnamefont {Gutt}},
  \bibinfo {author} {\bibfnamefont {L.~G.}\ \bibnamefont {Huang}}, \bibinfo
  {author} {\bibfnamefont {J.}~\bibnamefont {Metzkes}}, \bibinfo {author}
  {\bibfnamefont {U.}~\bibnamefont {Schramm}}, \bibinfo {author} {\bibfnamefont
  {M.}~\bibnamefont {Bussmann}}, \ and\ \bibinfo {author} {\bibfnamefont
  {T.~E.}\ \bibnamefont {Cowan}},\ }\bibfield  {title} {\enquote {\bibinfo
  {title} {{Using X-ray free-electron lasers for probing of complex interaction
  dynamics of ultra-intense lasers with solid matter}},}\ }\href {\doibase
  10.1063/1.4869331} {\bibfield  {journal} {\bibinfo  {journal} {Physics of
  Plasmas}\ }\textbf {\bibinfo {volume} {21}},\ \bibinfo {pages} {033110}
  (\bibinfo {year} {2014})}\BibitemShut {NoStop}%
\bibitem [{\citenamefont {Sherlock}\ \emph {et~al.}(2016)\citenamefont
  {Sherlock}, \citenamefont {Rozmus}, \citenamefont {Hill},\ and\ \citenamefont
  {Rose}}]{Sherlock2016}%
  \BibitemOpen
  \bibfield  {author} {\bibinfo {author} {\bibfnamefont {M.}~\bibnamefont
  {Sherlock}}, \bibinfo {author} {\bibfnamefont {W.}~\bibnamefont {Rozmus}},
  \bibinfo {author} {\bibfnamefont {E.~G.}\ \bibnamefont {Hill}}, \ and\
  \bibinfo {author} {\bibfnamefont {S.~J.}\ \bibnamefont {Rose}},\ }\bibfield
  {title} {\enquote {\bibinfo {title} {{Sherlock et al. Reply}},}\ }\href
  {\doibase 10.1103/PhysRevLett.116.159502} {\bibfield  {journal} {\bibinfo
  {journal} {Physical Review Letters}\ }\textbf {\bibinfo {volume} {116}},\
  \bibinfo {pages} {159502} (\bibinfo {year} {2016})}\BibitemShut {NoStop}%
\bibitem [{\citenamefont {Califano}\ \emph {et~al.}(1998)\citenamefont
  {Califano}, \citenamefont {Prandi}, \citenamefont {Pegoraro},\ and\
  \citenamefont {Bulanov}}]{Califano1998}%
  \BibitemOpen
  \bibfield  {author} {\bibinfo {author} {\bibfnamefont {F.}~\bibnamefont
  {Califano}}, \bibinfo {author} {\bibfnamefont {R.}~\bibnamefont {Prandi}},
  \bibinfo {author} {\bibfnamefont {F.}~\bibnamefont {Pegoraro}}, \ and\
  \bibinfo {author} {\bibfnamefont {S.~V.}\ \bibnamefont {Bulanov}},\
  }\bibfield  {title} {\enquote {\bibinfo {title} {{Nonlinear filamentation
  instability driven by an inhomogeneous current in a collisionless plasma}},}\
  }\href {\doibase 10.1103/PhysRevE.58.7837} {\bibfield  {journal} {\bibinfo
  {journal} {Physical Review E}\ }\textbf {\bibinfo {volume} {58}},\ \bibinfo
  {pages} {7837--7845} (\bibinfo {year} {1998})}\BibitemShut {NoStop}%
\bibitem [{\citenamefont {Silva}\ \emph {et~al.}(2002)\citenamefont {Silva},
  \citenamefont {Fonseca}, \citenamefont {Tonge}, \citenamefont {Mori},\ and\
  \citenamefont {Dawson}}]{Silva2002}%
  \BibitemOpen
  \bibfield  {author} {\bibinfo {author} {\bibfnamefont {LuÃ­s~O.}\
  \bibnamefont {Silva}}, \bibinfo {author} {\bibfnamefont {Ricardo~A.}\
  \bibnamefont {Fonseca}}, \bibinfo {author} {\bibfnamefont {John~W.}\
  \bibnamefont {Tonge}}, \bibinfo {author} {\bibfnamefont {Warren~B.}\
  \bibnamefont {Mori}}, \ and\ \bibinfo {author} {\bibfnamefont {John~M.}\
  \bibnamefont {Dawson}},\ }\bibfield  {title} {\enquote {\bibinfo {title} {{On
  the role of the purely transverse Weibel instability in fast ignitor
  scenarios}},}\ }\href {\doibase 10.1063/1.1476004} {\bibfield  {journal}
  {\bibinfo  {journal} {Physics of Plasmas}\ }\textbf {\bibinfo {volume} {9}},\
  \bibinfo {pages} {2458} (\bibinfo {year} {2002})}\BibitemShut {NoStop}%
\bibitem [{PIC(2017)}]{PIConGPU}%
  \BibitemOpen
  \href {http://picongpu.hzdr.de} {\enquote {\bibinfo {title} {{PIConGPU -- A
  Many-GPGPU Particle-in-Cell Code: http://picongpu.hzdr.de}},}\ } (\bibinfo
  {year} {2017})\BibitemShut {NoStop}%
\end{thebibliography}%
\end{document}